\documentclass[%
 reprint,
superscriptaddress,
 amsmath,amssymb,
 prl,
]{revtex4-2}

\usepackage{graphicx}
\usepackage{dcolumn}
\usepackage{bm}
\usepackage[margin=1in]{geometry}
\usepackage[italicdiff]{physics}
\usepackage{amsmath}
\usepackage{amssymb}
\usepackage{amsfonts}
\usepackage{algorithm}
\usepackage{algpseudocode}
\usepackage[most]{tcolorbox}
\usepackage{ragged2e}
\usepackage{mdframed}
\usepackage{appendix}
\usepackage{color}
\usepackage[colorlinks=true,allcolors=blue,breaklinks=true]{hyperref}
\usepackage{physics}
\usepackage[hints,insection]{minitoc}
\mtcsetdepth{secttoc}{2}  

\usepackage[dvipsnames]{xcolor}

\begin{document}
\dosecttoc  

\title{Flow, dynamics and active fracture in hydraulic multicellular systems}

\author{John D. Treado}
\affiliation{Cluster of Excellence Physics of Life, TU Dresden, 01062 Dresden, Germany}
\affiliation{Max Planck Institute for the Physics of Complex Systems, 01187 Dresden, Germany}

\author{Arthur Boutillon}
\affiliation{Cluster of Excellence Physics of Life, TU Dresden, 01062 Dresden, Germany}

\author{Frank Jülicher}\thanks{julicher@pks.mpg.de}
\affiliation{Cluster of Excellence Physics of Life, TU Dresden, 01062 Dresden, Germany}
\affiliation{Max Planck Institute for the Physics of Complex Systems, 01187 Dresden, Germany}
\affiliation{Center for Systems Biology Dresden, 01307 Dresden, Germany}

\author{Otger Campàs}\thanks{otger.campas@tu-dresden.de}
\affiliation{Cluster of Excellence Physics of Life, TU Dresden, 01062 Dresden, Germany}
\affiliation{Center for Systems Biology Dresden, 01307 Dresden, Germany}
\affiliation{Max Planck Institute of Molecular Cell Biology and Genetics, 01307 Dresden, Germany}

\date{}
\begin{abstract}
    From interstitial space to luminal cavities, fluid pressure and flow can remodel, reshape and even redefine a biological tissue. Fluids can either govern or react to mechanical interactions between cells. However, measuring flows at cellular scales is difficult, which makes it challenging to understand tissue hydraulics. Here, we develop a theoretical approach that captures cellular mechanics and fluid flow in one framework. We find that hydraulics can drastically influence tissue behavior. Hydraulic coupling between cell shape and size governs a tissue's response to osmotic shock, while tuning a tissue's permeabilities can channel fluid either between or across cell membranes. In active tissues, hydraulics can suppress cell mobility to the point of fracture, where we discover a hydraulic ratchet that drives fluid out of cells to generate small luminal spaces. We find experimental evidence that hydraulics can suppress cell motion in early stage zebrafish embryos injected with a thickening agent, which indicates that hydraulics may generally govern the behaviors of many multicellular systems. 
\end{abstract}

\maketitle

Tissue dynamics and behavior are constrained by their material properties. Properties like density, viscosity and elasticity derive from cell mechanics~\cite{Lecuit.Lenne.2007, Lenne.Viasnoff.2021}, cell-cell interactions~\cite{Campàs.Yap.2024, Ladoux.Mège.2017}, and interactions between cells and their extracellular matrix~\cite{Humphrey.Schwartz.2014, Helvert.Friedl.2018}. Recently, fluid pressure and flow have emerged as important mechanical elements in their own right. For example, the volume fraction of interstitial fluid can determine embryonic tissue rigidity~\cite{Mongera.Campàs.2018, Schliffka.Maître.2019, Petridou.Heisenberg.2019, Petridou.Hannezo.2021}. Exchange of cytoplasmic fluid between cells can drive instabilities and determine cell fate~\cite{Alsous.Martin.2021, Chartier.Grill.2021}. Tissues wounded in osmotic gradients can experience hydraulic fracture~\cite{Kennard.Theriot.2023}. Pressurized interstitial fluid can induce luminal cavity formation~\cite{Torres-Sánchez.Salbreux.2021, Schliffka.Maître.2024, Dagher.Maître.2024, Bovyn.Haas.2024}, and fully-formed lumina can exert pressures of kilopascals on surrounding cells~\cite{Dumortier.Maître.2019, Chan.Hiiragi.2019, Swinburne.Megason.2018, Lee.Grapin-Botton.2026}. From such observations, a picture has emerged of tissues as active poroelastic media~\cite{Casares.Trepat.2015, Shankar.Mahadevan.2024, Liu.Guo.2025} where cellular interactions and extracellular fluid couple and compete to determine a tissue's material properties. 

\begin{figure*}
    \centering
    \includegraphics[width=\linewidth]{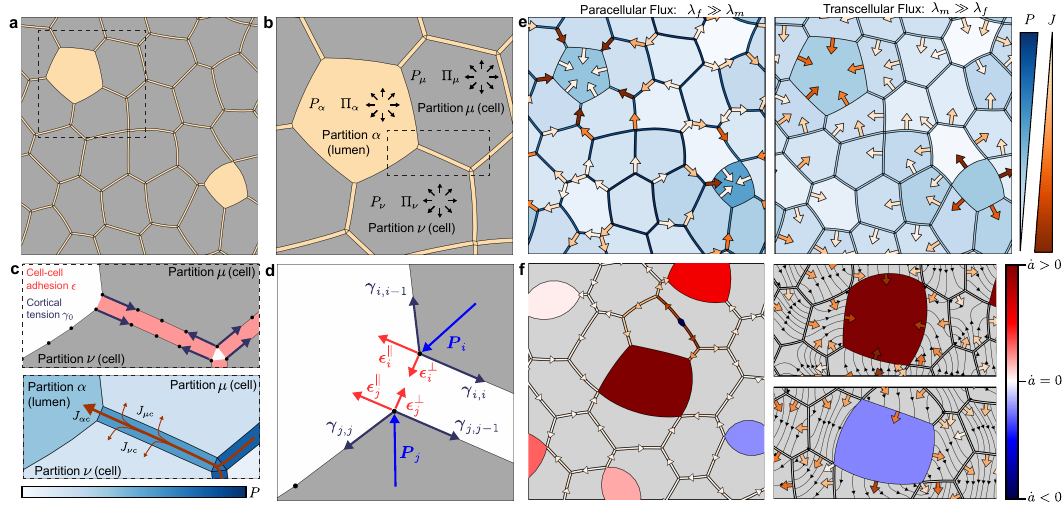}
    \caption{\textbf{Hydraulic partitions capture the coupling of cell mechanics and fluid flow} (a) A snapshot of a simulation of our model, where cells, channels (cell contact regions) and lumina (non-cell-contacting regions) are defined by polygonal partitions. (b) Each partition has a hydrostatic pressure $P$ and osmotic pressure $\Pi$. (c) Cell-cell adhesion with scale $\epsilon$ and cortical tension $\gamma_0$ (top panel), as well as fluid flow exchanged between connected partitions of different hydrostatic pressures $P$ (bottom panel), drive cell motion and fluid flux. (d) Forces (Eq.~\eqref{eq:mom_cons_main_eq}) applied to each vertex capture the features in (c). (e) Fluxes $J$ across cell interfaces (arrows, orange colorbar) are driven by differences in the hydrostatic pressure $P$ (blue colorbar) and by two permeabilities; $\lambda_m$, the permeability of cell membranes to flux, and $\lambda_f$, the permeability of fluid-fluid (i.e. channel-lumen) interfaces to flow. Varying the ratio $\lambda_f/\lambda_m$ tunes whether flux travels across cell interfaces (paracellular flux, left) or cell membranes (transcellular flux, right). (f) Flux direction drives partitions to either grow or shrink in time, with the rate of change of area $\dot{a}$ of lumina given by the colorbar. As in (e), flux arrows are colored with their flux magnitude $J$, and either drive paracellular (left) or transcellular (right) flux. In the transcellular case, we draw effective streamlines of fluid flow that would emerge from Stokes' flow constrained by the fluxes imposed at cell boundaries (see \hyperref[methods:streamlines]{Methods}). }
    \label{fig:Fig1}
\end{figure*}

Despite these advances, understanding how cell behavior impacts tissue hydraulics remains challenging. Only recently has osmotic pressure been measured \emph{in situ}~\cite{Vian.Campàs.2023}, and measurements of water flow at cellular resolution in living systems are lacking. Theory has played an important role in improving our understanding of how cells impact tissue material properties. Typical approaches are either cell-based, such as vertex models~\cite{Farhadifar.Jülicher.2007, Fletcher.Shvartsman.2014, Alt.Salbreux.2017, Popović.Wyart.2021} and phase-field models~\cite{Loewe.Marchetti.2020, Wang.Camley.2025}, or continuum theories~\cite{Duclut.Jülicher.2019, Duclut.Jülicher.2022, Kuan.Zaburdaev.2021, Merkel.Jülicher.2017, Etournay.Eaton.2015, Dye.Jülicher.2021}, which can capture both active and passive material properties of tissues as well as viscous and elastic behaviors. Vertex models have been key to understanding how cells influence epithelial mechanics, though combining such models with hydraulics is an open challenge. Recent work has extended vertex models to include extracellular space~\cite{Kim.Campàs.2021}. A next step is then to capture the physics of fluid flow across boundaries and along cell junctions.

In this work, we introduce a biophysical model of cell-fluid interactions in two-dimensional multicellular systems. We endow deformable polygons~\cite{Boromand.Shattuck.2018, Boromand.O'Hern.2019} with tensile surfaces and an adhesive interaction potential to capture the physics of contractile actomyosin cortices~\cite{Lecuit.Lenne.2007} and adherens junctions~\cite{Maître.Heisenberg.2012, Campàs.Yap.2024}. To capture the role of hydraulics, we incorporate incompressible fluid exchange between cells and the extracellular medium according to hydrostatic and osmotic pressure imbalance. By coupling cell mechanics with hydraulics, our model allows us to explore how flowing fluid impacts tissue material properties. We demonstrate how cell shape and size interplay during osmotic shock, how flow channeling emerges in tissues exposed to fluid pressure gradients, and how hydraulics can suppress cell mobility and even fracture cell-cell contacts in active tissues. We validate in part our model predictions with observations \emph{in vivo}, indicating that hydraulics can play a vital and general role in living tissue.

\subsection{Hydraulic partitions couple mechanics and flow}

We represent the geometry of two-dimensional tissues by partitioning space into cells, channels, and lumina (Fig.~\ref{fig:Fig1}a, Sec.~\ref{suppsec:hydraulic_tess_model}); channels are regions with adhesion-mediated contact between cells, while lumina fill the rest of space in which there is no cell-cell contact. With this representation, we can incorporate the physics of fluid exchange between partitions. 

Each partition has an osmotic pressure $\Pi$ and hydrostatic pressure $P$ (Fig.~\ref{fig:Fig1}b), differences of which can drive a flux $J$ that transports fluid between connected partitions (Fig.~\ref{fig:Fig1}c). The flux $J_{\mu\nu}$ between partitions $\mu$ and $\nu$ is
\begin{equation}\label{eq:flux_def}
    J_{\mu \nu} = \ell_{\mu\nu}\qty[j_{\mu\nu}+\lambda_{\mu\nu} (\Delta P_{\mu\nu} - \Delta \Pi_{\mu\nu})]
\end{equation}
where $j_{\mu\nu}$ is an active flux density resulting from the activity of ion pumps, $\Delta P_{\mu\nu} = P_\nu - P_\mu$ ($\Delta\Pi_{\mu\nu} = \Pi_\nu - \Pi_\mu$) is the hydrostatic (osmotic) pressure difference between partitions, respectively, and $\lambda_{\mu\nu}$ and $\ell_{\mu\nu}$ are the permeability and length of the interface separating the partitions. If fluid is incompressible (see Sec.~\ref{suppsec:hydraulic_tess_model:mass_cons}), the rate of change of area of each partition $\mu$ is governed by the net fluid flux entering the partition according to
\begin{equation}\label{eq:mass_cons_sum}
    \frac{d a_\mu}{dt} = \sum_{\nu } J_{\mu \nu}.
\end{equation}
In this work, we set the active flux density $j_{\mu\nu} = 0$.

Flux is thus governed by the balance of osmotic and hydrostatic pressure and regulated by interface permeability. We consider the area $a_c$ of each cell $c$ to directly determine cell osmotic pressure through
\begin{equation}
    \Pi_c = \frac{\kappa}{a_c - a_{0c}},
\end{equation}
where $\kappa$ and $a_{0c}$ represent the osmotic compressibility and dry area of trapped osmolytes trapped within cell $c$, respectively. We assume an external reservoir of osmolytes holds extracellular osmolytic concentrations fixed, enforcing a constant osmotic pressure $\Pi_0$ in any extracellular partition. 

\begin{table*}[t!]
    \centering
    \begin{tabular}{|l|l|}
        \hline
        Parameter & Description \\
        \hline
        $\tau_s / \tau_a$ & Ratio of shape to areal relaxation timescales ($\tau_s = \zeta_0 \sqrt{a_0} / \gamma_0$, $\tau_a = a_0^{3/2} / \lambda_m \kappa$) \\
        $\tau_\gamma / \tau_s$ & Ratio of tension persistence to shape relaxation timescales \\
        $\gamma_0 \sqrt{a_0} / \kappa$ & Ratio of Laplace pressure $\gamma_0/\sqrt{a_0}$ to osmotic pressure scale $\kappa/a_0$ \\
        $\epsilon / \gamma_0$ & Ratio of adhesion strength to cortical tension scale \\
        $\Pi_0 a_0 / \kappa$ & Ratio of extracellular osmotic pressure $\Pi_0$ to intracellular osmotic pressure scale $\kappa/a_0$ \\
        $\lambda_f / \lambda_m$ & Ratio of fluid to membrane permeability \\
        $\sigma_\gamma / \gamma_0$ & Strength of cortical tension fluctuation magnitude relative to the average cortical tension \\
        $\rho = N a_0 / A$ & Scaled cell number density in systems of domain area $A$\\
        \hline
    \end{tabular}
    \caption{Table of dimensionless parameters}
    \label{tab:params}
\end{table*}

For the permeability matrix $\lambda_{\mu\nu}$, we assign a constant membrane permeability $\lambda_m$ if either $\mu$ or $\nu$ is a cell partition; otherwise, we assign a constant fluid permeability $\lambda_f$. Cell membrane permeability $\lambda_m$ stems from aquaporins and the bare permeability of lipid bilayers, while fluid permeability $\lambda_f$ is set by both fluid viscosity and transmembrane proteins at cell interfaces that can resist flow.

While flux drives cell size changes according to Eq.~\eqref{eq:mass_cons_sum}, a fluid's hydrostatic pressure also exerts a force on cell membranes~\cite{Kennard.Theriot.2023, Dinet.Staykova.2023, Dumortier.Maître.2019, Casares.Trepat.2015, Gebala.Gerhardt.2016}. Furthermore, cell surface stresses like cortical tension~\cite{Lecuit.Lenne.2007} and adhesion~\cite{Maître.Heisenberg.2012} drive cell shape changes (Fig.~\ref{fig:Fig1}c). We apply these forces to vertices of every deformable polygon cell to drive cell movement and deformation. Assuming each vertex $i$ moves in an overdamped medium with friction coefficient $\zeta$, momentum conservation gives 
\begin{equation}\label{eq:mom_cons_main_eq}
    \zeta \frac{d \pmb{r}_i}{dt} = \vb*{\gamma}_i + \vb*{\epsilon}_i + \vb*{P}_i
\end{equation}
where $\vb*{r}_i$ is the position of the vertex and $\vb*{\gamma}_i$, $\vb*{\epsilon}_i$ and $\vb*{P}_i$ are the net forces due to tension, adhesion and hydrostatic pressure at vertex $i$, respectively (Fig.~\ref{fig:Fig1}d). The cortical tension force $\pmb{\gamma}_i$ arises from neighboring vertices via $\pmb{\gamma}_{ii}$ and $\pmb{\gamma}_{i,i-1}$ (Fig.~\ref{fig:Fig1}d), while the adhesion force $\vb*{\epsilon}_i$ can be decomposed into a contribution orthogonal (parallel) to interfacial channels $\vb*{\epsilon}_i^\perp$ ($\vb*{\epsilon}_i^\parallel$) (Fig.~\ref{fig:Fig1}d), and is controlled by an adhesion scale $\epsilon$. Definitions of the cortical tension and hydrostatic pressure forces are given in Sec.~\ref{suppsec:hydraulic_tess_model:mom_cons}, while the adhesion force is described in Sec.~\ref{suppsec:adhesion}.

\begin{figure*}[t!]
    \centering
    \includegraphics[width=\linewidth]{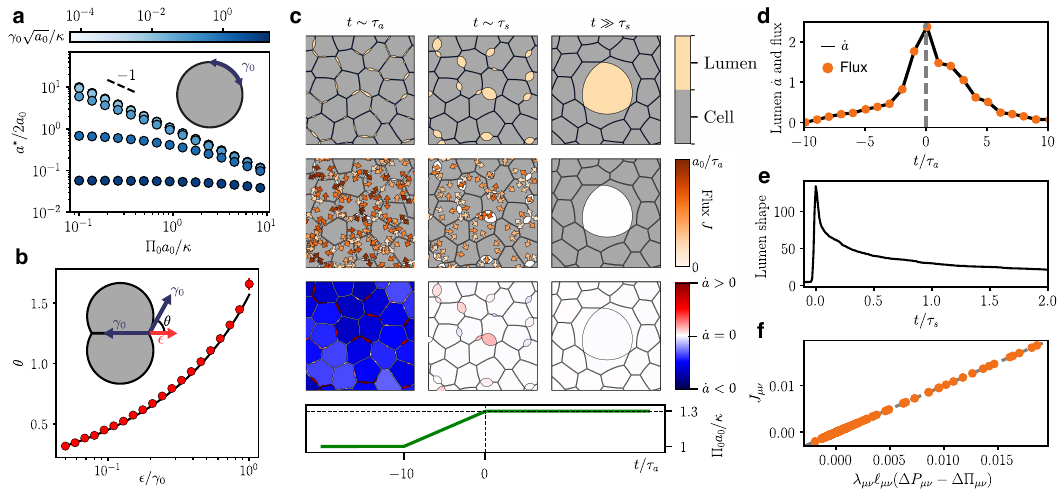}
    \caption{\textbf{Tissue fracture during osmotic shock} (a) Steady state area $a^*$ of a single cell for different values of the hydrostatic-to-osmotic pressure ratio $\gamma_0\sqrt{a_0}/\kappa$ as a function of external-to-internal osmolarity ratio $\Pi_0 \kappa / a_0$ (see Table~\ref{tab:params}). (b) Contact angle $\theta$ of cell doublets as a function of the adhesion-tension ratio $\epsilon/\gamma_0$. Black line denotes the  Young-Dupré law ($\cos\theta = 1 - \epsilon/\gamma_0$). (c) Osmotic shock in a multicellular tissue shown near times $\tau_a$, $\tau_s$ and $\gg \tau_s$ (see Table~\ref{tab:params}). Rows visualize partitions (top), flux magnitudes $J$ (middle) and rate of area change $\dot{a}$. For flux, arrows dictate direction and color dictates magnitude. Below, the scaled value of the external osmotic shock as a function of time is shown. The osmotic shock has an onset at $t = -10\tau_a$ and reaches a maximum at $t = 0$. (d) Net lumen flux and rate of area change $\dot{a}$ relax over timescales similar to $\tau_a$ and the osmotic shock ramp duration $10\tau_a$. (e) Lumen shape ($p_{\rm lumen}^2 / 4\pi a_{\rm lumen}$ for total lumen perimeter $p_{\rm lumen}$ and area $a_{\rm lumen}$) relaxes on timescales similar to $\tau_s$. (f) Comparison between all fluxes $J_{\mu\nu}$ between any two connected partitions $\mu$ and $\nu$ compared to the expected value from Eq.~\eqref{eq:flux_def}. In (c)-(e), $N=64$, $\rho = 0.5$, $\tau_s/\tau_a = 100$, $\epsilon/\gamma_0 = 0.5$, $\gamma_0\sqrt{a_0}/\kappa = 10^{-2}$, and $\lambda_f/\lambda_m = 1$. }
    \label{fig:Fig2}
\end{figure*}

Actomyosin cortex fluctuations can play a critical role in tissue dynamics and fluidization~\cite{Mongera.Campàs.2023, Kim.Campàs.2021, Curran.Baum.2017, Yanagida.Chalut.2022}. We therefore drive the interfacial tension $\gamma_{cc'}$ between cell $c$ and $c'$ with the modified Ornstein-Uhlenbeck process
\begin{equation}\label{eq:cortical_noise}
    \tau_\gamma \frac{d\gamma_{c c'}}{dt} = \gamma_0 - \gamma_{c c'} + \eta(t)\Theta(\gamma_{c c'} - \epsilon)
\end{equation}
where $\gamma_0$ is the steady-state value of cell cortical tension, $\tau_\gamma$ is the persistence timescale of active tension fluctuations, and $\eta$ is Gaussian white noise with correlation $\langle \eta(t)\eta(t')\rangle = \sigma_\gamma^2 \delta(t - t')$. The Heaviside step function $\Theta$ prevents negative effective tensions that lead to interfacial buckling.

Defining $\sqrt{a_0}$ as a characteristic length, $\gamma_0/\sqrt{a_0}$ as the cell Laplace pressure, and $\kappa/a_0$ as the intracellular osmotic pressure scale leads us to find two characteristic time scales in addition to $\tau_\gamma$; $\tau_a = a_0^{3/2} / \lambda_m\kappa$ and $\tau_s = \gamma_0 \zeta / \sqrt{a_0}$, which set the relaxation time of cell shape and area, respectively (\hyperref[methods:timescales]{Methods}). Scaling all times by $\tau_s$ then give all relevant parameters in the model, summarized in Table~\ref{tab:params}. In this work, we will focus on three hydraulic parameters: the ratio $\tau_s/\tau_a$;  $\Pi_0 a_0/\kappa$, which sets the balance of external to internal osmolarity scales; and $\lambda_f/\lambda_m$, which controls transcellular ($\lambda_f / \lambda_m \ll 1$) or paracellular ($\lambda_f / \lambda_m \gg 1$) flows (see Fig.~\ref{fig:Fig1}e). 

Combining the equations above allow us to determine the hydrostatic pressures $P$ of each partition as dynamic Lagrange multipliers (see \hyperref[methods:hydrostatic_pressure]{Methods}) that simultaneously enforce mass conservation (Eq.~\eqref{eq:mass_cons_sum}) and momentum conservation (Eq.~\eqref{eq:mom_cons_main_eq}). This approach provides a precise definition of the term ``hydraulics": the dual nature of fluid pressure to both exert forces on material surfaces and to drive flow between and across them. Numerical integration (\hyperref[methods:numerical_intergration]{Methods}) then provides the dynamics of the system. 

\subsection{Hydraulic coupling of cell shape and size}

By simultaneously enforcing both mass and momentum conservation, our model couples cell size with cell shape. While size is determined primarily by the balance of osmotic pressure in cells~\cite{Rollin.Sens.2023}, and shaped by cortical tension~\cite{Maître.Heisenberg.2012}, our model indicates that flow in multicellular contexts can be governed by both. We thus first interrogate the interplay of cell shape and size in our model. 

In steady state, the area $a^*$ of a single cell can be determined by the balance of hydrostatic and osmotic pressure. When the ratio $\gamma_0 \sqrt{a_0} / \kappa \ll 1$ (Table~\ref{tab:params}), $a^*$ is controlled exclusively by the external osmolarity (Fig.~\ref{fig:Fig2}a) where $a^* \sim \Pi_0^{-1}$ follows the 2D equivalent to Ponder's relation for cell volume~\cite{Ponder.Saslow.1931}. Increasing cortical tension increases the ratio of cell hydrostatic to osmotic pressure, and thus decreases cell area. Throughout the rest of this work, we fix the ratio $\gamma_0 \sqrt{a_0} / \kappa = 10^{-2}$ to capture the dominance of osmotic pressure on cell size regulation~\cite{Vian.Campàs.2023, Fischer-Friedrich.Helenius.2014, Rollin.Sens.2023, Venkova.Piel.2022}. 

The shape of adherent cells in steady state is determined by the ratio of cell adhesion $\epsilon$ to the scale of cortical tension $\gamma_0$~\cite{Maître.Heisenberg.2012}. Balancing the forces at a cell-cell interface, the contact angle $\theta$ of cell interfaces follows $\cos\theta = (\gamma_0 - \epsilon)/\gamma$, or the two-droplet extension of the Young-Dupré relation for droplet wetting (Fig.~\ref{fig:Fig2}b). Beyond two cells in mechanical equilibrium, varying $\epsilon/\gamma_0$ and dimensionless cell number density $\rho$ produce varied tissue morphologies (see Ref.~\cite{Kim.Campàs.2021}, Extended Data Fig.~\ref{fig:ExtDataFig_EqLimits}a) with $z=6$ average cell contacts in the confluent ($\rho\to \infty$) limit (Extended Data Fig.~\ref{fig:ExtDataFig_EqLimits}b). Tissue packing fraction $\phi$ track closely with $2\rho$ (Extended Data Fig.~\ref{fig:ExtDataFig_EqLimits}c), and cell shapes match those seen previously in densely packed simulated tissues~\cite{Bi.Manning.2015, Boromand.Shattuck.2018}. Interestingly, porous tissues exhibit a larger fraction of 4-sided cell junctions (Extended Data Fig.~\ref{fig:ExtDataFig_EqLimits}e), highlighting that tissue network topology emerges naturally in our model, possibly to stabilize tissues with highly non-spherical cells~\cite{Yan.Bi.2019}. 

While cell shape and size can be separated in the limit $\gamma_0\sqrt{a_0}/\kappa \ll 1$ in simple and equilibrium systems, non-equilibria can couple size and shape dynamically. To probe this coupling in multicellular systems, we expose an initially confluent tissue to an osmotic shock (Fig.~\ref{fig:Fig2}c). After a period of pure size response on timescales of $\tau_a$ (Fig.~\ref{fig:Fig2}d) with an increase in extracellular space, lumina between cells emerge on timescales of $\tau_s$ and continually round and coarsen over the course of the simulation. By tracking fluxes $J_{\mu\nu}$ over time, we verify that all fluxes follow the constitutive relation imposed in Eq.~\eqref{eq:flux_def} (Fig.~\ref{fig:Fig2}f). 

The effect of this osmotic shock is akin to hydraulic fracture observed in synthetic membranes~\cite{Dinet.Staykova.2023}, \emph{in vitro}~\cite{Casares.Trepat.2015}, and \emph{in vivo}~\cite{Dumortier.Maître.2019}, where excess fluid pressure in small regions near cell contacts drive the opening of small fluid pockets. Hydraulic fracture has been observed in tissues with osmotic gradients caused by wounding~\cite{Kennard.Theriot.2023}, indicating our model can capture this important feature in osmotically-stressed tissues. 

\begin{figure}[t]
    \centering
    \includegraphics[width=\linewidth]{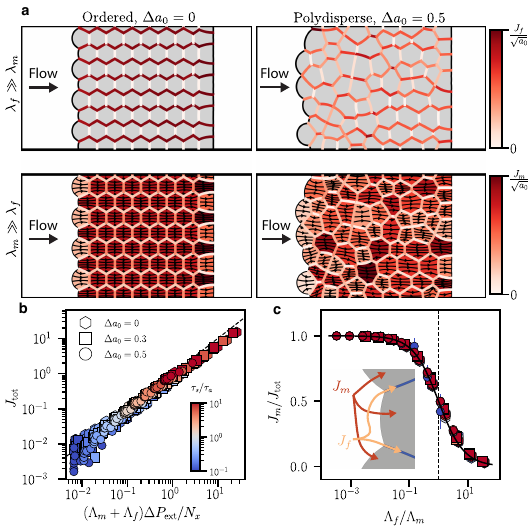}
    \caption{
        \textbf{Permeability ratio governs flow channeling in an external pressure gradient.} (a) Tuning the permeability ratio $\lambda_f/\lambda_m$ tunes fluid flux through fluid-filled channels ($\lambda_f \gg \lambda_m$, top row) or across cell membranes ($\lambda_f \ll \lambda_m$, bottom row) for both ordered tissues (left column, polydispersity $\Delta a_0 = 0$) or disordered tissues with high polydispersity (right column, polydispersity $\Delta a_0 = 0.5$) when an pressure drop $\Delta P_{\rm ext}$ is applied. Color denotes the transiting  flow velocity $v_t$ of the flow (see Main text). Streamlines denote the inferred direction of flux, as in Fig.~\ref{fig:Fig1}f. (b) Total flux $J_{\rm tot}$ through the tissue for a given pressure drop $\Delta P_{\rm ext} = 10\gamma_0/\sqrt{a_0}$ versus the bipartite network model flux $(\Lambda_f + \Lambda_m) \Delta P_{\rm ext}/N_x$ (see Eq.~\eqref{eq:simple_jtot} and \hyperref[methods:flow_model]{Methods}) for systems with varying polydispersity $\Delta a_0$ and timescale ratio $\tau_s/\tau_a$. The dashed line indicates Eq.~\eqref{eq:simple_jtot}. (c) Fraction of flux routed through cell membranes $J_m / J_{\rm tot}$ as a function of the hydraulic conductance ratio $\Lambda_f / \Lambda_m$ and the timescale ratio $\tau_s/\tau_a$. In both (b) and (c), colors indicate $\tau_s/\tau_a$ (colorbar) and symbols indicate polydispersity ratio (legend). Throughout, cells are initialized on a hexagonal lattice with $N_x = 8$, $N_y = 4$ (see \hyperref[methods:microfluidics]{Methods}), $\epsilon/\gamma_0 = 0.5$, $\Pi_0 a_0/\kappa = 1$, $\sigma_\gamma/\gamma_0 = 0$, and $\rho=0.35$. 
    }
    \label{fig:Fig3}
\end{figure}

\subsection{Tissue permeabilities channel fluid flow}

Beyond controlling tissue density, adhesion and osmolarity, our model allows us to study tissue response to externally-imposed flow. Regulation of trans- and paracellular flow plays a vital role in lumen inflation~\cite{Chan.Hiiragi.2019, Dumortier.Maître.2019, Mukenhirn.Honigmann.2024} and renal fluid reabsorption~\cite{Garcia.Knox.1998}; dysregulation of this balance can in turn lead to serious disease, such as vasogenic edema~\cite{Stokum.Simard.2015}. We therefore study flow across a simulated tissue generated by an imposed constant pressure ($\Delta P_{\rm ext} = 10 \gamma_0 / \sqrt{a_0}$). To vary the effect of tissue geometry on flow, we vary tissue polydispersity by drawing cell minimum areas from a normal distribution with mean $a_0$ and standard deviation $\sigma_a$, and randomize cell positions with a period of active crawling before relaxing cell positions and driving the tissue with external flow (\hyperref[methods:microfluidics]{Methods}). 

We quantify the rate at which fluid is transported across a given partition with the transiting flow velocity $v_t$ (see Eq.~\eqref{eq:methods:transit_flow_vel_def} in \hyperref[methods:flux_decomposition]{Methods}), which we obtain by decomposing the fluxes $J_{\mu\nu}$ (Eq.~\eqref{eq:flux_def}) into flux that either drives area change or that passes across the partition (\hyperref[methods:flux_decomposition]{Methods}). In both ordered and disordered tissues, the permeability ratio $\lambda_m/\lambda_f$ selectively guides flow through either channel or cell partitions (Fig.~\ref{fig:Fig3}a). While flow is uniform in ordered tissues, disorder in cell shape and size leads to disorder in the transit flow velocity. Disordered flow through disordered tissues is reminiscent of force chains in granular materials~\cite{Majmudar.Behringer.2005}, current paths in random resistor networks~\cite{Wu.Stanley.2005}, and flow paths in porous materials~\cite{Residori.Kurzthaler.2025}. Similar effects might allow tissues to tune disorder and membrane permeabilty in order to tune the path flows take across tissues. 

\begin{figure*}
    \centering
    \includegraphics[width=\linewidth]{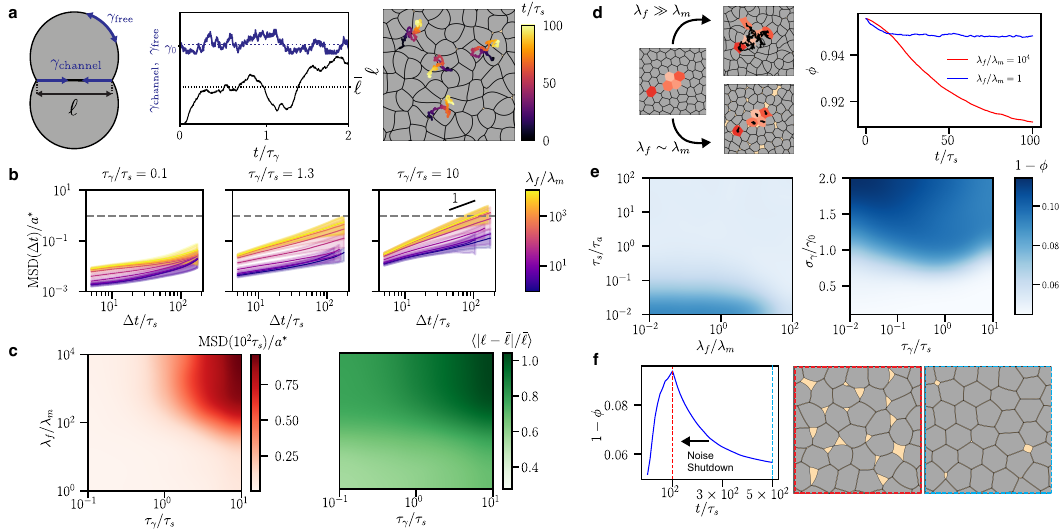}
    \caption{\textbf{Hydraulics govern cell mobility and can cause hydraulic fracture in active tissues}. (a) Noise is injected into tissues by driving the cortical tension at cell-channel interfaces ($\gamma_{\rm channel}$) and free cell interfaces ($\gamma_{\rm free}$) according to Eq.~\eqref{eq:cortical_noise}, which drives length fluctuations and the motion of cell centers. (b) Mean-square displacements (see \hyperref[methods:msd]{Methods}) of cell centers across different persistent times $\tau_\gamma/\tau_s$ (panels) and permeability ratios $\lambda_f/\lambda_m$ (colorbar). (c) MSD at $t = 10^2\tau_s$ (left) and time-averaged junctional strains $\langle \abs{\ell - \bar{\ell}}/\bar{\ell} \rangle$ (right, see \hyperref[methods:junctional_strain]{Methods}) across permeability ratio and persistence times. In (b) and (c), $\tau_s/\tau_a = 10^{-2}$, and $\sigma_\gamma/\gamma_0 = 2.5$. (d) Snapshots of simulations over time show that tissues with $\tau_s = 10^2\tau_a$ are dynamic with neighbor exchanges, while tissues with $\tau_s/\tau_a = 10^{-2}$ show less cell motion and the emergence of small lumina with decreasing cell packing fraction $\phi$ over time. (e) The luminal fraction $1 - \phi$ at steady-state across permeability ratios $\lambda_f/\lambda_m$ and timescale ratios $\tau_s/\tau_a$ (left), or persistence timescale ratios $\tau_\gamma/ \tau_s$ and scaled noise strengths $\sigma_\gamma/\gamma_0$ (right). At left, $\sigma_\gamma/\gamma_0 = 1$ and $\tau_\gamma/\tau_s = 1$, whereas at right $\lambda_f/\lambda_m = 1$ and $\tau_s/\tau_a = 10^{-2}$. (f) Example of luminal fraction $1 - \phi$ over time when noise is removed at $t = 10^2 \tau_s$. Throughout, $\rho = 0.5$, $\epsilon/\gamma_0=0.5$, and $N=64$ except in panel (f), where $N=24$. }
    \label{fig:Fig4}
\end{figure*}

We can calculate the total flux $J_{\rm tot}$ exiting the tissue using a simple bipartite flow network (\hyperref[methods:flow_model]{Methods}, Extended Data Fig.~\ref{fig:ExtDataFig_Bipartite}). From the permeabilities $\lambda_m$ and $\lambda_f$, we define the hydraulic conductances $\Lambda_m = \ell_m \lambda_m$ and $\Lambda_f = \ell_f \lambda_f$, where $\ell_m$ and $\ell_f$ are the interfacial lengths cell membrane or fluid partitions with the right-most fluid chamber, respectively. In agreement with the bipartite network calculation, the total flux of fluid through a simulated tissue $J_{\rm tot}$ is well described by the effective flux
\begin{equation}\label{eq:simple_jtot}
    J_{\rm tot} = \frac{\Lambda_m + \Lambda_f}{N_x} \Delta P_{\rm ext},
\end{equation}
where $N_x$ is the number of cells initially placed in the horizontal (i.e. flow) direction in the hexagonal lattice. As shown in Fig.~\ref{fig:Fig3}b, this relation holds across permeabilities, relaxation timescale ratios and levels or tissue disorder. We further show that flux is controlled by the conductance ratio $\Lambda_m / \Lambda_f$; the fraction of flow across membranes $J_m/J_{\rm tot}$ displays sigmoidal behavior as a function of $\Lambda_m / \Lambda_f$, with flux splitting evenly between fluids and cell membranes when $\Lambda_m = \Lambda_f$ (Fig.~\ref{fig:Fig3}c).  

\subsection{Tissue hydraulics and activity can suppress cell mobility and fracture tissues}

\begin{figure*}
    \centering
    \includegraphics[width=\linewidth]{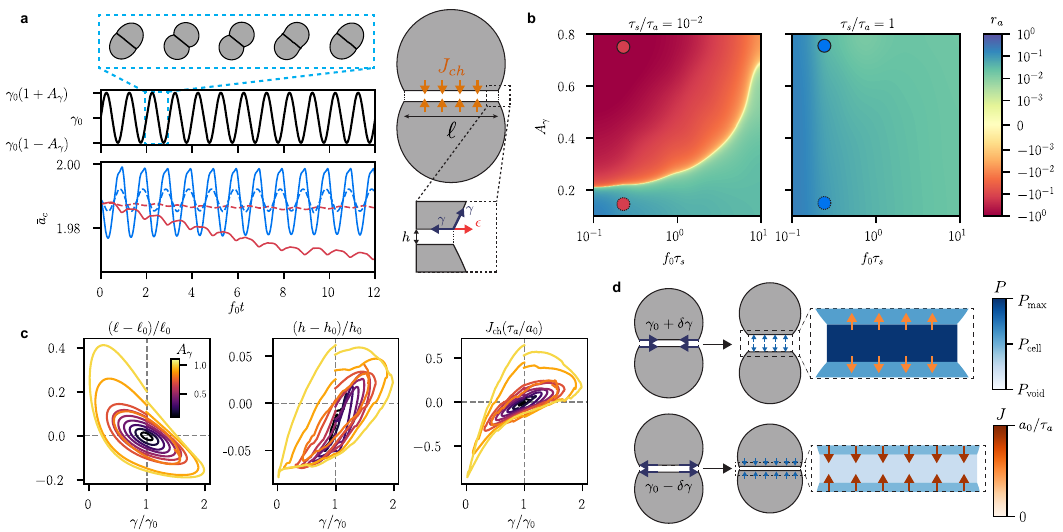}
    \caption{\textbf{Oscillating cortical tension drives a hydraulic ratchet} (a) Schematic of oscillatory cortical tension between to adhesive cell doubles. Mean cell area $\bar{a}_c$ for $\tau_s/\tau_\gamma = 10^{-2}$ ($=1$) shown in blue (red) for both low driving amplitude $A_\gamma=0.1$ (dashed lines) and high driving amplitude $A_\gamma=0.9$ (solid lines) (see Eq.~\eqref{eq:osc_cort_tens}). During the simulation, we track the interface length $\ell$, the interfacial separation $h$ and net flux from the cell to the interfacial channel $J_{\rm ch}$. (b) Cell area response $r_a$, which measures whether cells lose ($r_a<0$) or gain ($r_a>0$) area over successive oscillatory cycles relative to the amplitude of driving (see \hyperref[methods:cell_area_response]{Methods}) for $\tau_s/\tau_\gamma = 10^{-2}$ (left) and $=1$ (right). (c) Phase-space orbits of either the interfacial length $\ell$, height $h$, and channel-cell flux $J_{\rm ch}$ versus the scaled cortical tension $\gamma(t)/\gamma_0$, averaged over the first 4 oscillatory periods for simulations with $\tau_s/\tau_a = 10^{-2}$ and $f_0 \tau_s = 10^{-1}$. The interfacial length (height) is relative to the initial length value $\ell_0$ ($h_0$), flux $J_{\rm ch}$ is relative to the flux scale $a_0/\tau_a$, and color denotes the driving amplitude $A_\gamma$ (see colorbar). (d) Schematic of the hydraulic ratchet mechanism that drives fluid out of cells when $\tau_s/\tau_a \ll 1$. Throughout this figure, $\lambda_f/\lambda_m=1$, $\Pi_0 a_0/\kappa = 1$, and $\epsilon/\gamma_0 = 0.5$.}
    \label{fig:Fig5}
\end{figure*}

Cell motion due to activity can also drive fluids to flow within tissues. However, it is unclear whether these flows then feedback into tissue dynamics. In principle, fluid flow generated by cell activity could either hinder or increase cell motility. To investigate the effect of cell actvity on fluid flow, we simulate dense active tissues with $\sigma_\gamma/\gamma_0 > 0$ (Eq.~\eqref{eq:cortical_noise}) and vary both the permeability ratio $\lambda_f/\lambda_m$ and the persistence of cortical fluctuations. For persistence times $\tau_\gamma > \tau_s$, cell center mean-square displacements (MSD, see \hyperref[methods:msd]{Methods}) show diffusive behavior and exceed the steady state cell area $a^*$ after sufficient run times (Fig.~\ref{fig:Fig4}b), indicating the presence of neighbor exchange and tissue fluidity~\cite{Stooke-Vaughan2025}. 

\begin{figure}[h!]
    \centering
    \includegraphics[width=\linewidth]{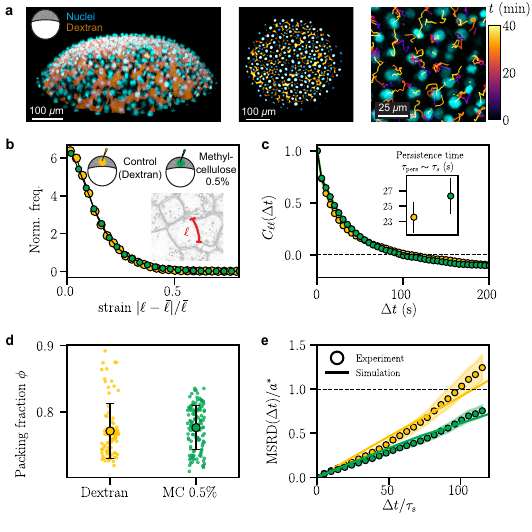}
    \caption{\textbf{Methylcellulose injection suppresses cell motion \emph{in vivo}} (a) Confocal microscopy images of a zebrafish blastoderm in the sphere stage, showing nuclear markers (cyan) and extracellular space (fluorescent Dextran, yellow). Nuclei are tracked over time (right, colorbar). (b) Normalized frequency (area under curves $= 1$) of junctional strain $\abs{\ell - \bar{\ell}}/\bar{\ell}$ (\hyperref[methods:embryo_juncs]{Methods}) observed for junctions of length $\ell$ (inset). Data taken from $n=633$ ($n=498$) junctions observed in the Dextran (Methylcellulose) conditions. (c) Temporal autocorrelation $C_{\ell \ell}(\Delta t)$ of junctional length fluctuations (\hyperref[methods:embryo_juncs]{Methods}), with measured persistence time $\tau_{\rm pers}$ reported in the inset. (d) Packing fraction measured in plane (panel a, \hyperref[methods:embryo_packing_fraction]{Methods}). Data taken from $n=6$ ($n=8$) embryos in the Dextran (Methylcellulose) conditions, with a total of $85$ ($124$) planes analyzed. (e) Mean-square relative displacement (MSRD, \hyperref[methods:MSRD]{Methods}) scaled by cell size $a^*$ of nuclear tracks (circles) from embryos injected with either Dextran (control, yellow) or Methylcellulose 0.5\% solution (green). Simulation parameters are described in main text. To compare simulation and experiment, we scale the MSRD time lag by $\tau_s$ from the simulation, and $\tau_{\rm pers}$ measured in experiments. Throughout, green (yellow) indicates data from embryos injected with Dextran (Methylcellulose 0.5\% solution by weight). }
    \label{fig:Fig6}
\end{figure}

However, cell-center MSD decreases upon decreasing $\lambda_m/\lambda_f$, regardless of the persistence of cortical fluctuations (Fig.~\ref{fig:Fig4}b). Comparing MSDs at a fixed time $10^{2}\tau_s$ reveals significant suppression of cell mobility even with highly persistent cortical tension fluctuations (Fig.~\ref{fig:Fig4}b, c), and little change to the average junctional strain $\langle \abs{\ell - \bar{\ell}}/\bar{\ell} \rangle$ (Fig.~\ref{fig:Fig4}c; \hyperref[methods:junctional_strain]{Methods}). This suggests a purely hydraulic origin of cell mobility suppression, where low fluid permeability restricts the fluid flow between cell interfaces necessary for cell neighbor exchange. 

In this limit, tissue structure changes significantly. Small lumina emerge at cell-cell contacts and grow over time, leading to a decrease in cell packing (Fig.~\ref{fig:Fig4}d). While spontaneous lumina also appeared during osmotic shock (Fig.~\ref{fig:Fig2}), here, endogenous activity rather than externally-controlled osmolarity drives the emergence of luminal space. We find fracture occurs when $\tau_s \ll \tau_a$, $\lambda_f \leq \lambda_m$, and when noise strength $\sigma_\gamma /\gamma_0$ is sufficiently large (Fig.~\ref{fig:Fig4}e). Lumina generated by fracture do not coalesce beyond a certain size (Extended Data Fig.~\ref{fig:ExtDataFig_NoVoidCoalescence}), and fluid is reabsorbed into cells when activity is removed (Fig.~\ref{fig:Fig4}f). 

While cell activity is known to fracture contacts by virtue of cell motion~\cite{Lv.Li.2024, Wang.Camley.2026, Sonam.Ladoux.2023, Prakash.Prakash.2021} or growth~\cite{Treado.O’Hern.2022}, our model exhibits an active form of hydraulic fracture due to increased fluid pressure of fluid between cells. Ratchets, i.e. driven systems with directional flow, have been observed and characterized in a variety of active systems~\cite{Granek.Solon.2024}. To understand if and how a ratchet emerges in our model, we simulate adhesive cell doublets driven by oscillatory cortical tension $\gamma(t)$ about an average cortical tension $\gamma_0$ (Fig.~\ref{fig:Fig5}a), where
\begin{equation}\label{eq:osc_cort_tens}
    \gamma(t) = \gamma_0\qty[1+ A_\gamma\cos(2\pi f_0 t)],
\end{equation}
and $f_0$ ($A_\gamma$) is the driving frequency (driving amplitude normalized by $\gamma_0$). We focus on the regime $\lambda_f < \lambda_m$ where hydraulic fracture is possible (Fig.~\ref{fig:Fig4}e) and study if oscillating cortical tension in this simple system is sufficient to drive fluid out of cells and into the surrounding space. 

We quantify the scale of cell size response to driving by computing the strain-to-driving ratio $r_a$ (see \hyperref[methods:cell_area_response]{Methods}), which compares the ratio between the relative drop in mean cell area $\bar{a}_c$ to the dimensionless scale of max cortical tension $A_\gamma \gamma_0 \sqrt{a_0} / \kappa$. When $\tau_s/\tau_a \ll 1$, cell sizes decrease over time when $A_\gamma$ exceeds an $f_0$-dependent threshold (Fig.~\ref{fig:Fig5}b). 

To understand this mechanism in more detail, we analyze flux between cells and the interstitial channel ($J_{\rm ch}$), the interface length $\ell$ and the distance between cell surfaces $h$ during the transient relaxation of the cell doublet system (Fig.~\ref{fig:Fig6}c). When cortical tension decreases, channels expand to preserve volume, leading to a drop in the channel pressure and flow outward from cells into the channel. The reverse occurs when the cortical tension increases, although the flux back into cells is lower (Fig.~\ref{fig:Fig5}d). Flux is proportional to length (Eq.~\eqref{eq:flux_def}), so longer interfaces at lower cortical tension transport more fluid out of the cell than during the second half of the cycle. 

This effect constitutes a hydraulic ratchet and can account for the emergence of small lumina in active multicellular systems; net fluid flow out of cells can lead to an accumulation of fluid at cell-cell junctions, increase pressure, and eventually fracture the contact. Because this process is generic and independent of biological details, we predict that decreased fluid permeability suppresses cell motion and generates luminal nucleation and growth in active tissues.

\subsection{Cell motion is suppressed in zebrafish embryos injected with methylcellulose}

To test these predictions, we investigated the dynamics of cells in the dynamic and porous zebrafish blastoderm at the sphere stage. To label the extracellular fluid and control its viscosity, we injected fluorescent Dextran either alone or in a 0.5\% methylcellulose solution (MC) into the extracellular space (\hyperref[methods:zfish]{Methods}). Though the Dextran solution has viscosity similar to water, we measured that the dynamic viscosity of 0.5\% MC was roughly 6$\times$ larger (see \hyperref[methods:dextran_vs_methly]{Methods}). This value is at least 5 orders of magnitude smaller than tissue viscosity measured in the sphere stage~\cite{Petridou.Heisenberg.2019}, and 6 orders of magnitude smaller than those in the zebrafish tailbud~\cite{Mongera.Campàs.2023}. Thus, any change to tissue dynamics should derive from purely hydraulic, rather than frictional, effects. 

If MC injection were to simply increase friction, it should primarily do so at cell junctions. We first segmented and analyzed the fluctuations of cell junctional lengths $\ell$ by measuring the distribution of junctional strain ($\abs{\ell - \bar{\ell}}/\bar{\ell}$) and the temporal autocorrelation of lengths ($C_{\ell \ell}$; \hyperref[methods:embryo_juncs]{Methods}). We find no difference in either metric between our two experimental conditions (Fig.~\ref{fig:Fig6}b,c), indicating that MC presence in the extracellular space does not significantly alter junction fluctuations. This echoes our previous result that junctional strain in simulations does not significantly depend on the permeability of the extracellular medium (Fig.~\ref{fig:Fig4}c). 

We next measure the tissue packing fraction $\phi$ in order to see if MC injection affects tissue structure, which could also affect cell mobility by affecting tissue mechanics~\cite{Mongera.Campàs.2018}. However, we find no change to tissue packing fraction in either experimental condition (Fig.~\ref{fig:Fig6}d). This observation contradicts our prediction that decreased fluid permeability drives hydraulic fracture in active tissues. To understand this discrepancy, we performed simulations at packing fractions similar to the experiment (Extended Data Fig.~\ref{fig:ExtDataFig_MSDComp}), and find that tissues with $\phi=0.8$ do not display hydraulic fracture, while denser tissues with $\phi \geq 0.9$ do. This simulation result suggests that the zebrafish sphere stage blastoderm lacks sufficient cell packing for cell activity to drive hydraulic fracture. 

However, upon nuclei tracking and measurement of the mean-square relative displacement (MSRD; \hyperref[methods:MSRD]{Methods}), we find that MC injection does lead to a lower MSRD compared to Dextran injection. To compare experiments with simulations in similar conditions, we simulate tissues prepared at similar packing fractions ($\phi=0.76$), high noise ($\sigma_\gamma/\gamma_0 = 5$) and lower adhesion ($\epsilon/\gamma_0 = 0.15$) to prevent clustering. We note that we cannot independently constrain our hydraulic parameters, as absolute measurements of permeability and $\tau_a$ in living embryos are lacking. We find simulations at relaxation timescale ratio $\tau_s/\tau_a \approx 0.5$ whose MSDs match the experimental conditions. We find that $\lambda_f/\lambda_m \approx 8$ ($\approx 2$) correspond well to MSDs observed in embryos injected with Dextran (MC). The ratio between fluid permeabilities in these two simulations corresponds to $\approx 4$, close to the viscosity ratio observed between Dextran and MC ($\approx 6$), indicating that the simulations qualitatively reproduce the behavior observed in experiments. 

\subsection{Discussion}

Our work reveals that hydraulic coupling between cells and fluid can guide flows through tissues, modify cell mobility and alter a tissue's architecture. The generality of our model indicates that hydraulics should play an important role beyond luminal or vascular tissues. Indeed, we find that hydraulics affects tissue dynamics in the zebrafish sphere stage blastoderm, a system with no organs or obvious fluid flows. Our results therefore call for future experimental studies on the hydraulics of other non-luminal tissues, such as epithelial monolayers or porous mesenchyme, to further investigate the extent to which hydraulics impact tissue behavior. Our work so far does not study the effect of directed fluid transport, which can play a important role in lumenogenesis~\cite{Dumortier.Maître.2019}. Future work will investigate how active transport can modify the results shown here. This will broaden our understanding of the role of hydraulics in tissues and organs. 

\subsection{Acknowledgements}

The authors thank Matthew Bovyn, Matteo Ciarchi, Omer Granek, Lara Koehler, Marko Popović, Vincenzo Maria Schimmenti, and Amit Singh Vishen for useful discussions. 

\subsection{Code and Data Availability}

Data and simulation code will be made available after the review process.

\bibliography{Hydraulics_of_Active_Tissues, Arthur_Refs, repo}

\pagebreak

\clearpage
\setcounter{equation}{0}
\setcounter{secnumdepth}{2} 
\renewcommand{\thesection}{M\arabic{section}}
\renewcommand{\thesubsection}{M\arabic{subsection}}
\renewcommand{\theequation}{M\arabic{equation}}

\section*{Methods}\label{sec:methods}

\subsection{Hydrostatic Pressure Determination} \label{methods:hydrostatic_pressure}

The rate of change of area of a particular partition $\mu$ (cell, channel or lumen) is determined by the sum of total fluxes $J_{\mu\nu}$ from other partitions $\nu$ (Eq.~\eqref{eq:mass_cons_sum}). By the chain rule, this is also given by
\begin{equation}
    \dv{a_\mu}{t} = \sum_{\text{verts }i} \frac{d a_\mu}{d \vb*{r}_i}\dv{\vb*{r}_i}{t}.
\end{equation}
The vertex velocity $\dv*{\vb*{r}_i}{t}$ is determined by momentum conservation in Eq.~\eqref{eq:mom_cons_main_eq}, which means we can derive a closed expression for the hydrostatic pressures given forces applied to each vertex, osmotic balance and external pumping. In Supplemental Material Section~\ref{suppsec:hydraulic_tess_model}, we show this closed equation can be expressed in the form
\begin{equation}\label{eq:methods:lin_pressure_eq}
    A \vb{P} = \vb{\Omega}
\end{equation}
where $\vb{P}$ is the hydrostatic pressures in all partitions, $A$ is a square symmetric matrix with units of area, and $\vb{\Omega}$ is a vector with units of energy that details the total energy required to change the area of a particular partition. By solving Eq.~\eqref{eq:methods:lin_pressure_eq}, we can determine the pressure in each partition. 

\subsection{Streamline Computation from Constrained Stokes Flow}\label{methods:streamlines}

In Fig.~\ref{fig:Fig1}a, we show streamlines through cells due to the presence of fluxes $J$ at cell boundaries. We draw the streamlines of an effective flow field $\pmb{u}$ due to each flux assuming the tissue domain is a single incompressible fluid with a uniform viscosity $\eta$ at low Reynold's number. The flow is thus governed by Stoke's equations
\begin{subequations}
	\begin{align}
		\eta \nabla^2 \pmb{u} &= \nabla p\\
		\nabla \cdot \pmb{u} &= 0
	\end{align}
\end{subequations}
where $p$ is the effective hydrostatic pressure field that enforces the incompressibility condition. Introducing the scalar stream function $\psi$ that satisfies $\pmb{u} = \nabla^\perp \psi = \nabla \times (\psi \hat{z})$~\cite{Tong.Tong.2025} allows us to solve for $\pmb{u}$ directly without solving for the pressure field; the contours of $\psi$ then give the streamlines we plot in Fig.~\ref{fig:Fig1}. To simplify our visualization, we solve for $\psi$ within each cell individually with the flux conditions at the boundaries as boundary conditions. Because we are only interested in the streamline direction and not their magnitude, we take $\eta = 1$. 

To implement the fluxes $J_e$ on each polygonal edge $e$ as a boundary condition for the flow $\pmb{u}$, we enforce that the flow $\pmb{u}$ satisfies
\begin{equation}
	J_e = -\int_e \pmb{u} \cdot \pmb{\hat{n}}_e \ ds
\end{equation}
where the integral is taken over the line defined by the edge $e$, and $\pmb{\hat{n}}_e$ is the vector pointing normal and outward from the cell with edge $e$. 

Let $\pmb{\hat{t}}_e$ be the unit vector tangent to edge $e$, such that $\pmb{\hat{n}}_e = (\hat{t}_y, -\hat{t}_x)$. Thus, $\pmb{u} \cdot \pmb{\hat{n}}_e = u_x \hat{t}_y - u_y \hat{t}_x$. Note that, for a planar curve $\pmb{r}(s) = \qty[x(s), y(s)]$ parameterized by a coordinate $s$, a function $f$ defined on the curve has the property $\partial f / \partial s = (\partial f/ \partial x) \hat{t}_x + (\partial f/ \partial x) \hat{t}_y$. Given the definition of $\pmb{u} = \qty(\partial \psi / \partial y, -\partial \psi /\partial x)$, we have $\pmb{u} \cdot \pmb{n}_e = \partial \psi / \partial s$ for the coordinate $s$ along the polygonal edge $e$, which implies, for an edge $e$ defined by the endpoints $\pmb{r}_A$ and $\pmb{r}_B$, we have
\begin{equation}
	\psi(\pmb{r}_B) - \psi(\pmb{r}_A) = \int_e\pmb{u} \cdot \pmb{\hat{n}}_e \ ds = -J_e. 
\end{equation}
Therefore, the stream function allows us to turn an integral constraint on $\pmb{u}$ into a Dirichlet boundary condition for $\psi$ evaluated at the vertices of the polygon within which we solve for the flow. We arbitrarily set $\psi = 0$ for one vertex, as only relative values of $\psi$ matter for the determination of streamlines within a polygon. Letting $\pmb{u} = \nabla^\perp \psi$, we obtain the biharmonic equation $\nabla^4 \psi = 0$ in the interior of the polygon~\cite{Tong.Tong.2025}. We also impose a no-slip boundary condition $\pmb{u} \cdot \pmb{\hat{t}}_e = 0$ for all edges $e$ on the polygon surface, which can be written as the von Neumann boundary condition $\nabla \psi \cdot \pmb{\hat{n}}_e = 0$ for the stream function. We then solve the biharmonic equation $\nabla^4 \psi = 0$ numerically using a finite-element, triangulated mesh generated within each polygon.

\subsection{Numerical integration} \label{methods:numerical_intergration}

We numerically integrate Eq.~\eqref{eq:mom_cons_main_eq} using the Boost odeint libary~\cite{s3} and the Cash-Karp adaptive Runge-Kutta algorithm to control for solution stiffness~\cite{Cash.Karp.1990}. The equation for cortical tension noise (Eq.~\eqref{eq:cortical_noise}) is numerically integrate with the Euler-Maryuama algorithm, though only when steps from the Cash-Karp algorithm are accepted. 

\subsection{Timescales \& dimensionless parameters }\label{methods:timescales}
In this section, we derive the governing timescales and dimensionless parameters shown in Table~\ref{tab:params}. As the minimum cell size $a_0$ scales with the cell size, we use $\sqrt{a_0}$ as the characteristic length scale. All vertex positions can be expressed as $\vb*{r}_i = \tilde{\vb*{r}}_i\sqrt{a_0}$ and all compartment areas can be expressed as $a_q = a_0 \tilde{a}_q$, where quantities with $\tilde{}$ are taken to be dimensionless. Time is then expressed as $t = \tau \tilde{t}$, where $\tau$ is a characteristic timescale. 

We then rescale lengths and times in the governing equations of motion of our model (Eqs.~\eqref{eq:mass_cons_sum},~\eqref{eq:mom_cons_main_eq} and~\eqref{eq:cortical_noise}) and obtain the following rescaled equations of motion
\begin{subequations}\label{eq:nondimEqMot1}
    \begin{align}
        \frac{\zeta \sqrt{a_0}}{\tau}\frac{d \tilde{\vb*{r}}_i}{d\tilde{t}} &= \vb*{\gamma}_i + \vb*{\epsilon}_i + \vb*{P}_i \label{subeq:force_eom_rescale}\\
        \frac{\sqrt{a_0}}{\tau}\frac{d \tilde{a}_\mu}{d\tilde{t}} &= \lambda_m\qty[\Delta P - \qty(\Pi_0-\frac{\kappa}{a_\mu-a_0})]\tilde{p}_\mu\label{subeq:area_eom_rescaled}\\
        \frac{\tau_\gamma}{\tau} \frac{d\tilde{\gamma}_{\mu \nu}}{d\tilde{t}} &= \qty[1 - \tilde{\gamma}_{\mu \nu} + \frac{\eta(t)}{\gamma_0}]\Theta(\gamma_{\mu \nu} - \epsilon) \label{subeq:noise_eom_rescaled}
    \end{align}
\end{subequations}
where we use in Eq.~\eqref{subeq:area_eom_rescaled} the mass conservation for a single cell with area $a_\mu = a_0\tilde{a}_\mu$ and perimeter $p_\mu = \tilde{p}_\mu\sqrt{a_0}$, and where we have immediately rescaled the interfacial cortical tensions $\gamma_{\mu \nu}$ between cells $\mu$ and $\nu$ by the cortical tension fixed point $\gamma_0$ in Eq.~\eqref{subeq:noise_eom_rescaled}. 

To non-dimensionalize forces, we rescale the adhesion force by the adhesion scale $\epsilon$ and the pressure force by the force scale $P^*\sqrt{a_0}$, where $P^*$ is a typical scale of hydrostatic pressure. Momentum conservation at each vertex $i$ can then be rewritten
\begin{equation}
    \frac{d \tilde{\vb*{r}}_i}{d\tilde{t}} = \qty(\frac{\tau \gamma_0}{\zeta \sqrt{a_0}})\qty[\tilde{\vb*{\gamma}}_i + \qty(\frac{\epsilon}{\gamma_0})\tilde{\vb*{\epsilon}}_i + \qty(\frac{P^*\sqrt{a_0}}{\gamma_0})\tilde{\vb*{P}}_i]
\end{equation}
We then identify the characteristic time scale of shape change
\begin{equation}
    \tau = \tau_s = \frac{\zeta\sqrt{a_0}}{\gamma_0}
\end{equation}
as the time required for cortical tension to move a vertex a distance $\sqrt{a_0}$ against friction. We also identify the hydrostatic pressure scale
\begin{equation}
    P^* = \gamma_0/\sqrt{a_0}
\end{equation}
which is the Laplace pressure for a circular cell with tension $\gamma_0$ and area $a_0$. The ratio $\epsilon/\gamma_0$ then controls the ratio between adhesion and cortical tension, which we show in Fig.~\ref{fig:Fig2}b to control the two-cell wetting contact angle $\theta$. 

We then analyze the time scales relevant for mass conservation. From Eq.~\eqref{subeq:area_eom_rescaled}, we rescale $\Delta P$ by the pressure scale $P^* = \gamma_0/\sqrt{a_0}$, and we rescale the osmotic pressure difference by
\begin{equation}
    \Pi^* = \frac{\kappa}{a_0},
\end{equation}
the typical scale of osmotic pressure in an individual cell. We arrive at the non-dimensionalized equation
\begin{equation}
    \frac{d \tilde{\vb*{r}}_i}{d\tilde{t}} = \frac{\tau_s}{\tau_a} \qty[\qty(\frac{P^*}{\Pi^*})\Delta\tilde{P} - \qty(\frac{\Pi_0}{\Pi^*} - \frac{1}{\tilde{a} - 1})]
\end{equation}
where we identify the characteristic time scale of area change
\begin{equation}
    \tau_a = \frac{a_0^{3/2}}{\lambda_m \kappa}
\end{equation}
as well as two other dimensionless quantities: the ratio $P^*/\Pi^*$, which describes the competition between hydrostatic and osmotic pressure in a cell, and $\Pi_0/\Pi^*$, which describes the competition between external and internal osmolarity. Similar analysis to Eq.~\eqref{subeq:noise_eom_rescaled} reveals that the characteristic time scale of noise relaxation is simply the persistence time scale $\tau_\gamma$, and the ratio $\sigma_\gamma / \gamma_0$ controls the relative deviations of fluctuating cortical tensions relative to their mean. 


\subsection{Microfluidic Channel Simulations}\label{methods:microfluidics} 

To simulate tissues with externally-posed pressure (Fig.~\ref{fig:Fig3}), we simulate a microfluidic channel periodic in the vertical ($y$) direction with two open chambers on either side of the tissue in the horizontal ($x$) direction. Cells are first seeded in a central channel with periodic boundary conditions in the vertical direction in a hexagonal lattice where all cells are given initial areas $a = 2a_0$. We place $N_x$ unit cells of the hexagonal lattice along the $x$ axis and $N_y$ unit cells along the $y$ direction, giving a total of $N = 2N_x N_y$ cells total. Disorder is introduced by drawing the minimum area of each cell $a_{0\mu}$ from a Gaussian distribution with mean $1$ and standard deviation $\Delta a_0$. If polydispersity is introduced, cell centers are mixed randomly by applying a self-propulsion force with scale $\zeta v_0$ to cell centers. The self-propulsion force acts along a director $\hat{n}_\mu = (\cos\theta_\mu, \sin\theta_\mu)$ with an angle $\theta_\mu$ that diffuses as $d\theta_\mu/dt = D_r \eta_\theta(t)$ with $\eta_\theta$ a Gaussian white noise with $0$ mean and correlation $\langle \eta_\theta(t)\eta_\theta(t')\rangle =\delta(t - t')$. This produces athermal Active Brownian dynamics~\cite{Fily.Marchetti.2012}, which is sufficient to mix cells and produce disordered packings. Cells are then brought to mechanical equilibrium by allowing all forces to relax.

Once in mechanical equilibrium, an external pressure gradient $\Delta P_{\rm ext}$ is imposed by fixing the pressure in the left-hand chamber in the simulated microfluidic channel. To restrict tissue translation with the fluid flux, a soft linear force $f_{\rm wall} = -k_{\rm wall}\delta$ is imposed in the $x$-direction on any vertex whose overlap $\delta$ with the wall is non-zero. We take $k_{\rm wall} = 100(\kappa/a_0)$ to minimize overlap of the tissue with the wall. To impose a fixed pressure in our fluid network, we modify the definition of the constraint vector $\vb{\Omega}$ (see Eq.~\eqref{eq:methods:lin_pressure_eq} and Sec.~\ref{suppsec:hydraulic_tess_model:constrain_sat}). For any partition $\mu$ that shares an interface of length $\ell_{\mu, {\rm ext}}$ with the left-hand chamber, the constraint vector for the microfluidic system $\Omega^{\rm ext}$ is given as
\begin{equation}
    \Omega_\mu^{\rm ext} = \Omega_\mu + \lambda_\mu \ell_{\mu, {\rm ext}}P_{\rm ext}
\end{equation}
where $\Omega_\mu$ is the constraint vector without constant pressure in the left-land chamber, and $\lambda_\mu$ is the appropriate permeability for the interface; either $\lambda_m$ if partition $\mu$ is a cell, or $\lambda_f$ if partition $\mu$ is a channel. Then, all pressures in the tissue (cells + fluid partitions) are determined by the modified equation $A\vb{P} = \vb{\Omega}^{\rm ext}$. 

\subsection{Flux decomposition and transit velocity}\label{methods:flux_decomposition}
To measure the rate at which fluid passes across a partition, we decompose the flux $J_{\mu\nu}$ into parts
\begin{equation}\label{eq:methods:flux_decomposition}
    J_{\mu\nu} = J_{\mu\nu}^{\rm bulk} + J_{\mu\nu}^{\rm transit}
\end{equation}
where $J_{\mu\nu}^{\rm bulk}$ captures the part of flux that changes cell area, and $J_{\mu\nu}^{\rm transit}$ captures flux that transits through the partition. As the net flux into cells drives area change (Eq.~\eqref{eq:mom_cons_main_eq}), we use a decomposition where $\sum_\nu J_{\mu\nu}^{\rm bulk} = da_\mu/dt$, and $\sum_\nu J_{\mu\nu}^{\rm transit} = 0$.  

We first define the fluid potential $\psi_{\mu} = P_\mu - \Pi_\mu$ of partition $\mu$ and the interface conductance $\Lambda_{\mu\nu}=\ell_{\mu\nu}\lambda_{\mu\nu}$ between partitions $\mu$ and $\nu$. Then, we can rewrite Eq.~\eqref{eq:flux_def} as
\begin{equation}
    J_{\mu\nu} = \Lambda_{\mu\nu}(\psi_\nu-\psi_\mu).
\end{equation}
Since $\sum_\nu J_{\mu\nu}^{\rm bulk} = da_\mu/dt$, the net bulk flux into a cell matches the total net flux, which allows us to define
\begin{equation}
    J_{\mu\nu}^{\rm bulk} = \Lambda_{\mu\nu}(\bar{\psi}_\mu - \psi_\mu)
\end{equation}
where 
\begin{equation}
    \bar{\psi}_\mu = \frac{\sum_{\nu}\Lambda_{\mu\nu} \psi_\nu}{\sum_\nu \Lambda_{\mu\nu}}
\end{equation}
is the conductance-weighted average of neighboring partitions' fluid potential. Eq.~\eqref{eq:methods:flux_decomposition} then gives
\begin{equation}
    J_{\mu\nu}^{\rm transit} = \Lambda_{\mu\nu}(\psi_\nu - \bar{\psi}_\mu).
\end{equation}
As the net transit flux must sum to $0$, $J_{\mu\nu}^{\rm transit}$ is either positive or negative depending on $\nu$. We define the transiting flow velocity $v_{t,\mu}$ of partition $\mu$ as the sum of positive transit fluxes scaled by the interfaces across which they are active. That is
\begin{equation}\label{eq:methods:transit_flow_vel_def}
    v_{t,\mu} = \frac{\sum_{\nu}J_{\mu\nu}^{\rm transit}\Theta(J_{\mu\nu}^{\rm transit})}{\sum_{\nu}\ell_{\mu\nu}\Theta(J_{\mu\nu}^{\rm transit})}
\end{equation}
where $\Theta$ is the Heaviside step function. 

\subsection{Bipartite network flow model}\label{methods:flow_model}

To calculate the flux through a tissue in the microfluidic channel depicted in Fig.~\ref{fig:Fig3}, we construct a simplified hydraulic network that contains the same salient features. We coarse-grain the ordered tissue as a network of cell-nodes and fluid-nodes, where all cell-nodes must connect to fluid-nodes, while fluid-nodes can connect to each other. Imposing periodic boundary conditions in the $y$-direction, the simplest such network where cells do not connect to the same channel across the periodic boundary is shown in Extended Data Fig~\ref{fig:ExtDataFig_Bipartite}. A source node held at constant pressure $P_{\rm ext}$ and sink node held at constant $0$ pressure are placed at the left and right boundaries of the network, respectively, and connect to all nodes at the edge of the tissue graph. Flux between connected nodes $i$ and $j$ is given by $J_{ij} = \Lambda_{ij}\Delta P_{ij}$, where $\Delta P_{ij} = P_j - P_i$ ($\Lambda_{ij}$) is the pressure difference (conductance) between nodes. Cell-fluid (fluid-fluid) connections are given a constant conductance $\Lambda_m$ ($\Lambda_f$). 

The graph has either rows with cells or rows without cells. Thus, by symmetry, the pressure of a node at depth (i.e. column) $j$ depends only on if the row contains cells or not. Let $P_c(j)$ ($P_f(j)$) be the pressure of a node in a row with (without) cells at depth $j$ into the tissue graph. Let $\Delta_{c,f}(j) = P_{c,f}(j+1) + P_{c,f}(j-1) - 2P_{c,f}(j)$ be the discrete Laplacian operator along either a cell ($c$) or fluid ($f$) rows, and let $\nabla(j) = P_f(j) - P_c(j)$ be the discrete gradient operator for flow between rows. Mass conservation at each node in steady state gives
\begin{subequations}
    \begin{align}
        \Lambda_m \Delta_c(j)+2\Lambda_m \nabla(j) &= 0 \ \forall j \text{ odd}\\
        \Lambda_m \Delta_c(j)+2\Lambda_f \nabla(j) &= 0 \ \forall j \text{ even}\\
        \Lambda_f \Delta_f(j)-2\Lambda_m \nabla(j) &= 0 \ \forall j \text{ odd}\\
        \Lambda_f \Delta_f(j)-2\Lambda_f \nabla(j) &= 0 \ \forall j \text{ even}
    \end{align}
\end{subequations}
Note that even columns contain cell nodes, whereas odd columns only contain fluid nodes (see Extended Data Fig.~\ref{fig:ExtDataFig_Bipartite}). By substitution, we find that $\nabla(j) = 0$, which means pressures in each node depend only on their tissue depth and not on the row in which the node is contained. Thus, the pressure $P(j)$ in any node at depth $j$ in the tissue is given by
\begin{equation}
    P(j+1)+P(j-1)-2P(j) = 0
\end{equation}
This discrete Laplace equation has the general linear solution $P(j) = A+Bj$. For a linear graph of $M/2$ interior fluid columns and $M/2 + 1$ cell columns, boundary conditions give $P(1)+P_{\rm ext}-2P(0) = P(M-1)-2P(M)=0$ and allow us to identify the pressure in a node at depth $j$ as
\begin{equation}
    P(j) = \frac{M+1-j}{M+1}P_{\rm ext}. 
\end{equation}
Thus, the flux into the final (exterior) node at depth $j = M + 1$ is given by flux from two cell nodes with flux $J_m =  2\Lambda_m P_{\rm ext}/(M+1)$ and from two fluid nodes with flux $J_f 2\Lambda_f P_{\rm ext} / (M+1)$. For a tissue with $N$ cells per row, the total number of columns is $M = 2N-2$. Thus, the total flux $J_{\rm tot} = J_m + J_f$ is given by
\begin{equation}\label{eq:JtotModel}
    J_{\rm tot} = \qty(\frac{\Lambda_m + \Lambda_f}{N})P_{\rm ext}
\end{equation}
and the ratio of cell flux to total flux is
\begin{equation}\label{eq:JratModel}
    \frac{J_m}{J_{\rm tot}} = \frac{1}{1+(\Lambda_f/\Lambda_m)}.
\end{equation}
Eq.~\eqref{eq:JtotModel} is the total flux that collapses data at high conductance in Fig.~\ref{fig:Fig3}b, and Eq.~\eqref{eq:JtotModel} is the sigmoidal function shown in Fig.~\ref{fig:Fig3}c.

\subsection{Mean-square displacement (MSD)}\label{methods:msd}

The MSD of a simulated tissue (Fig.~\ref{fig:Fig4}b,c) is defined as
\begin{equation}
   \text{MSD}(\Delta t) = \langle \qty(\vb*{R}_\mu(t+\tau) - \vb*{R}_\mu(t))^2\rangle
\end{equation}
$\vb*{R}_\mu(t)$ is the geometric center of cell $\mu$ at time $t$, and where brackets $\langle \rangle$ indicate averages over all cells $\mu$ and time origins $t$. In practice, for simulations of length $T$ with frames spaced in time by $dt$ that occur at times $t_k = k \ dt$, we define
\begin{equation}\label{eq:msd}
   \text{MSD}(\Delta t) = dt\sum_{\mu = 1}^N \sum_{k = 1}^{\frac{T-\Delta t}{dt}} \frac{\qty[\vb*{R}_\mu(t_k+\Delta t) - \vb*{R}_\mu(t_k)]^2}{N(T-\Delta t)}
\end{equation}
where the time delay $\Delta t$ is restricted to times $t_k = k\Delta t$ for $k = 1, ..., (T/dt) - 1$. 

\subsection{Junctional strain analysis}\label{methods:junctional_strain}

In Fig.~\ref{fig:Fig4}c, we introduce the average junctional strain $\langle \abs{\ell - \bar{\ell}}/\bar{\ell}\rangle$, which quantifies the extent to which cellular junctions deform relative their average length $\bar{\ell}$ over the course of their existence. To compute this quantity in a simulation, we define $\ell_{\mu\nu}(t)$ as the total junctional length between cell $\mu$ and $\nu$. We define this junctional as the sum of the length of extracellular channels that are shared by both cell $\mu$ and $\nu$; if cells $\mu$ and $\nu$ share no channels, then $\ell_{\mu\nu} = 0$.The quantity $\langle \abs{\ell - \bar{\ell}}/\bar{\ell}\rangle$ is then averaged over all junctions $\ell_{\mu\nu}$ from all pairs of cells $\mu$ and $\nu$ that share a channel in steady state. 

\subsection{Cell area response to oscillatory strain}\label{methods:cell_area_response}

To measure how much cell areas deform in response to oscillatory strain (Fig.~\ref{fig:Fig5}) with amplitude $\gamma_0(1 + A_\gamma)$, we define
\begin{equation}
    r_a = \frac{\delta \bar{a}_c }{\Sigma}
\end{equation}
where, for a simulation of duration $T$, we define the areal strain as
\begin{equation}
    \delta \bar{a}_c = \frac{\bar{a}_c(T) - \bar{a}_c(0)}{\bar{a}_c(0)}
\end{equation}
where $\bar{a}_c(t)$ is the mean cell area at time $t$ in the simulation. We define the scaled area stress $\Sigma$ as
\begin{equation}
    \Sigma = A_\gamma \qty(\frac{\gamma_0\sqrt{a_0}}{\kappa})
\end{equation}
which maps oscillations of scaled amplitude $A_\gamma$ in cortical tension to stress ratio between surface and bulk stresses. 

\subsection{Zebrafish embryos: handling, staging and injection}\label{methods:zfish} 
  
Zebrafish (\textit{Danio rerio}) were maintained as previously described~\cite{westerfield2000zebrafish}. Experiments were performed in accordance with all relevant ethical regulations, following protocols approved under European Union Directive 2010/63/EU and the German Animal Welfare Act. Because the sex of zebrafish embryos and larvae cannot be distinguished at the stages studied, sex-specific experiments were not required. For nuclear and membrane labelling we used \textit{Tg(h2afz:GFP)}\textsuperscript{kca6} \cite{pauls2001} and \textit{Tg(actb2:MA-mCherry2)}\textsuperscript{hm29}~\cite{Xiong2013}, respectively. 
  
Embryos were staged according to \cite{Kimmel1995}. At the oblong stage, 4nl of a solution containing 1mg/ml Rhodamine~B--dextran (MW 10\,000; Thermo Fisher), either alone or supplemented with 0.5\% methylcellulose (036718.36, Thermo Fisher) in distilled water, was microinjected into the extracellular space. After 10 minutes, the dextran signal was homogeneous throughout the extracellular space, and embryos were mounted in E3 medium containing 0.2\% agarose, animal pole facing the coverslip of a glass-bottom MatTek dish (Part No.\ P35G-0.170-14-C). 
  
\subsection{Confocal microscopy and image analysis}\label{methods:microscopy} 
  
Embryos with labelled nuclei were imaged on a confocal laser-scanning microscope (either a Zeiss LSM~980 or a Leica SP8) using a Plan-Apochromat $25\times$/0.8~W (Zeiss) or an HCX IRApo L $25\times$/0.95~W (Leica) objective, respectively. Imaging used two-photon excitation, with an InSight STDS-AX or an InSight DeepSee Dual laser (Spectra-Physics), tuned to 920 nm, which excites both GFP and rhodamine. Stacks of the inner cell mass were acquired every 2 minutes for 2 hours, with a 2 µm $z$-step. 
  
Embryos with labelled membranes were imaged on a confocal laser-scanning microscope (either a Zeiss LSM~700 or a Zeiss LSM~980) using a Plan-Apochromat $25\times$/0.8~W (Zeiss) objective. Single planes of cell membranes within the inner cell mass were acquired every 5 seconds for 10 minutes. 
  
All imaging was performed at 25°C. 
  
\subsection{Embryonic packing fraction}\label{methods:embryo_packing_fraction} 
  
To quantify the experimental packing fraction $\phi$, the rhodamine channel of nuclei-labelled embryos was extracted in Fiji. A cuboidal region of interest (ROI) spanning the depth of the embryo was cropped and, for each stack, the extracellular space was segmented using the Fiji Thresholding tool (Triangle method) to generate a binary mask. The packing fraction was then quantified as the percentage of null (zero-valued) pixels in each plane. 
  
\subsection{Viscosity of dextran versus methylcellulose solutions}\label{methods:dextran_vs_methly} 
  
The viscosities of dextran alone and of dextran with 0.5\% methylcellulose were measured on an MCR302e rheometer (Anton Paar, Germany) at 20°C $\pm\ 0.1 $K for shear rates from 0.01 to 100 s\textsuperscript{-1}. 

Between 1 - 100 s\textsuperscript{-1} shear rates, we measured constant viscosities of 0.91 mPa s$^{-1}$ (dextran alone) and 5.84 mPa s$^{-1}$ (dextran with methylcellulose). A second set of measurements, performed with an SV-10 vibro-viscometer (Malvern Instrument) at 30 Hz, yielded  0.88 mPa s$^{-1}$ and 5.03 mPa s$^{-1}$, respectively.

\subsection{Nuclear segmentation and tracking}\label{methods:segmentation} 
  
Imaging data were first pre-processed in Fiji. Inner cell mass confocal 3D time-lapses were smoothed with a 1-pixel median filter, and photobleaching was corrected using the Exponential Fit function. 
  
The fluorescently labelled inner cell mass nuclei were then segmented using the Spot Detection module of Imaris (version 11.0.1, Bitplane) and tracked with the Brownian-motion tracking algorithm. Tracks were visually inspected for validation. 
  
Because of cell division and developmental progression, cell diameter decreases over time. To stage-match embryos precisely for accurate comparison, the average nearest-neighbor distance was quantified over time as a proxy for cell diameter. For each embryo, we retained a 50 minutes window centered on the time at which the average cell diameter reached 25 $\mu$m, corresponding to cell diameters of 28 to 22 $\mu$m. 
  
\subsection{Mean-square relative displacement (MSRD)}\label{methods:MSRD} 
  
Cell trajectories were computed by tracking the center positions of all nuclei at each time point. To quantify cell movements, we computed the mean-square relative displacement (MSRD) as previously described~\cite{Mongera.Campàs.2018, Stooke-Vaughan2025}. The MSRD is defined similarly to the MSD (see Eq.~\eqref{eq:msd}), but neighbor information is taken into account to correct for drift and better reproduce the motion of cell centers relative to their local environment. We define the cell-cell distance between cell nuclear centers $\mu$ and $\nu$:
\begin{equation}
    \pmb{R}_{\mu\nu}(t) = \pmb{R}_\nu(t) - \pmb{R}_\mu(t)
\end{equation}
The MSRD is defined then as the mean square-displacement of $\pmb{R}_{\mu\nu}$ averaged over all neighbors $\nu$ of cell $\mu$ and scaled by the average cell size $\ell_c$. That is, we define
\begin{equation}
   \text{MSRD}(\Delta t) =  \frac{\langle |\pmb{R}_{\mu\nu}(t+\Delta t) - \pmb{R}_{\mu\nu}(t)|^2 \rangle}{\ell_c^2}
\end{equation}
where the average $\langle \rangle$ denotes an average over all cells $\mu$, the set of neighbors $\nu$ of all $\mu$ at time $t$, and over time origins $t$. We considered here a unique characteristic cell size $\ell_c$ of 25 $\mu$m. In simulations with no net drift of the center of mass of the tissue, the MSRD is equivalent to the MSD scaled by $\ell_c^2$; thus, to compare simulations and experiment in Fig.~\ref{fig:Fig6}b, we report simulate MSDs scaled by $\ell_c^2 = a^*$. where $a^*$ is the steady state area of the cells. 
  
\subsection{Embryonic junctional length measurement, analysis and correlations}\label{methods:embryo_juncs} 
  
Junctional lengths and their dynamics were determined as previously reported~\cite{Mongera.Campàs.2018, Mongera.Campàs.2023}. To monitor cellular junctions over time, we acquired confocal sections of membrane-labelled embryos every 5 seconds for a total of 10 minutes. Cell contours were obtained using Cellpose and custom Python code, followed by manual correction with the Tissue Analyzer plugin of Fiji. We selected all cell--cell contacts persisting for at least 120 s and computed their time-dependent normalized lengths $(\ell-\overline{\ell})/\overline{\ell}$, where $\ell$ is the cell--cell contact length at a given time point and $\overline{\ell}$ is the average length of that individual contact over time. The values of $(\ell-\overline{\ell})/\overline{\ell}$ for all time points and all cell--cell contacts were then combined into a single frequency distribution of the absolute normalized length $\lvert(\ell-\overline{\ell})/\overline{\ell}\rvert$ (Fig.~\ref{fig:Fig6}d). For a junction with temporal trace $\ell(t)$, we compute the autocorrelation $C_{\ell \ell}(\Delta t)$ at lag time $\Delta t$ using
\begin{equation}
    C_{\ell \ell}(\Delta t) = \frac{\langle \ell(t) \ell(t + \tau)\rangle - \langle\ell(t)\rangle^2}{\sqrt{\langle\ell^2(t)\rangle-\langle \ell(t) \rangle^2}}
\end{equation}
where the angle brackets $\langle \rangle$ denote averages over the time origin $t$. This quantity is scaled between $[-1, 1]$, and the correlation function reported in Fig.~\ref{fig:Fig6}e is averaged over the ensemble of junctional lengths observed in a particular experimental condition. 

\newpage
\pagebreak

\setcounter{figure}{0}
\renewcommand{\thefigure}{E\arabic{figure}}
\onecolumngrid
\begin{figure*}[h!]
    \centering
    \includegraphics[width=0.95\linewidth]{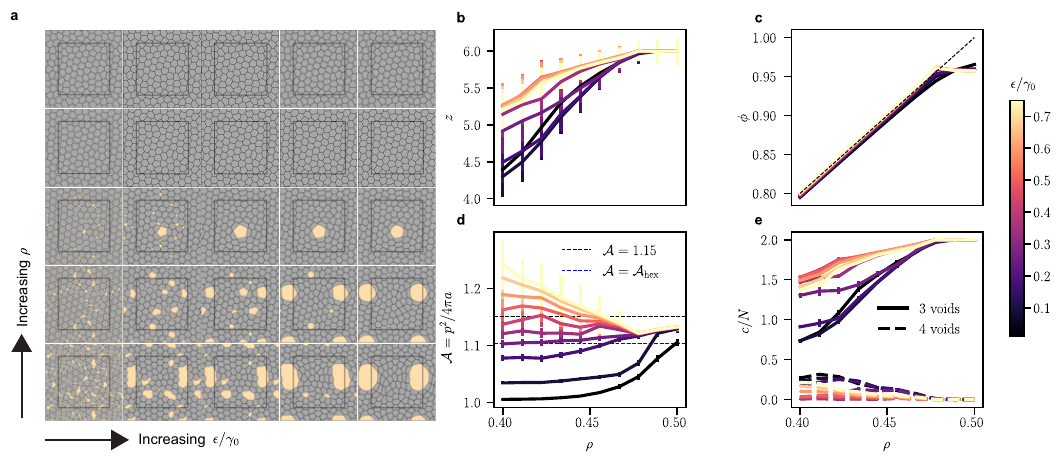}
    \caption{\textbf{Equilibrium behavior of multicellular simulations with hydraulic partitions}. (a) Example snapshots taken of the model in mechanical equilibrium as a function of dimensionless number density $\rho$ (rows) and scaled adhesion $\epsilon/\gamma_0$ (columns). In (b)-(e), we show average number of contacts $z$ (b), packing fraction $\phi$ (c), average cell shape parameter $\mathcal{A} = p^2/4\pi a$ for cells with perimeter $p$ and area $a$, compared to the shape parameter of a perfect hexagon $\mathcal{A}_{\rm hex} \approx 1.1$ and the expected value at confluence $\mathcal{A}=1.15$~\cite{Boromand.Shattuck.2018} (d), and the fraction of the coordination $c$ of small lumina at cell-cell junctions, either 3-sided (solid) or 4-sided (dashed) lumina (e). Throughout, $N=64$, and the relative adhesion $\epsilon/\gamma_0$ is given by the colorbar. In (c), the dashed line denotes $\phi = 2 \rho$; steady state areas in the limit $\gamma_0\sqrt{a_0}/\kappa \ll 1$ are close to $2a_0$ (Fig.~\ref{fig:Fig2}a), so $\phi$ should be near $2\rho$ in dilute packings. }
    \label{fig:ExtDataFig_EqLimits}
\end{figure*}

\pagebreak

\begin{figure*}[h!]
    \centering
    \includegraphics[width=\linewidth]{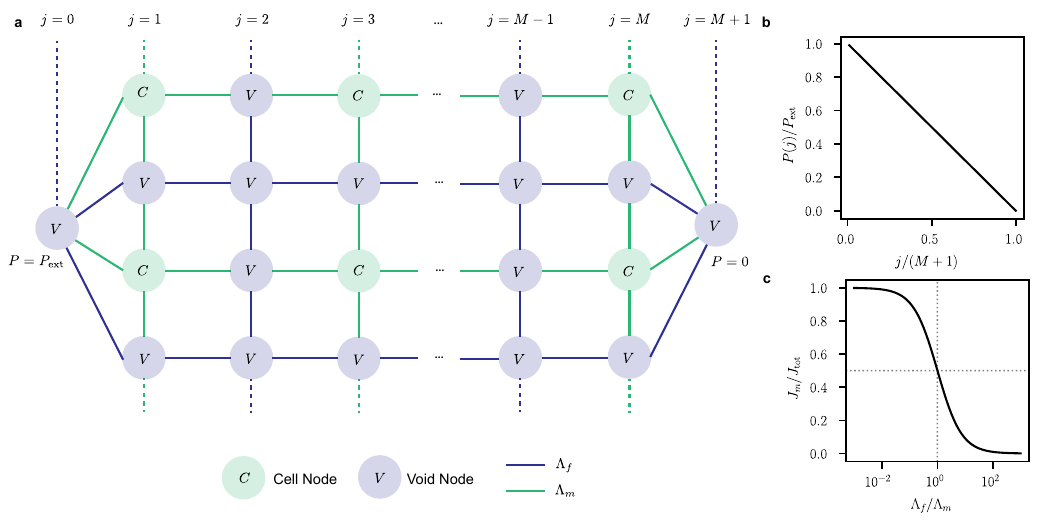}
    \caption{\textbf{Bipartite network model of microfluidic tissue flow explains net flux}.  (a) Bipartite network model of a tissue in a microfluidic chamber as described in \hyperref[methods:flow_model]{Methods} with $M$ interior nodes connected to $2$ exterior nodes held at constant pressure. The tissue network contains either cell ($C$, green) or void ($V$, blue) node. Cell and void nodes exchange fluid with conductivity $\Lambda_m$ and connected void nodes exchange fluid with conductivity $\Lambda_f$. Exterior void nodes are held at either $P_{\rm ext}$ and $0$ to maintain a pressure gradient across the tissue network. (b) Scaled pressure $P(j)/P_{\rm ext}$ at an interior node index $j$ as a function of the depth into the network. (c) Total flux entering cell membranes $J_m$ at the first interior nodes at $j=1$ scaled by the total flux $J_{\rm tot}$ as a function of the conductivity ratio $\Lambda_f / \Lambda_m$. }
    \label{fig:ExtDataFig_Bipartite}
\end{figure*}

\pagebreak

\begin{figure*}[h!]
    \centering
    \includegraphics[width=\linewidth]{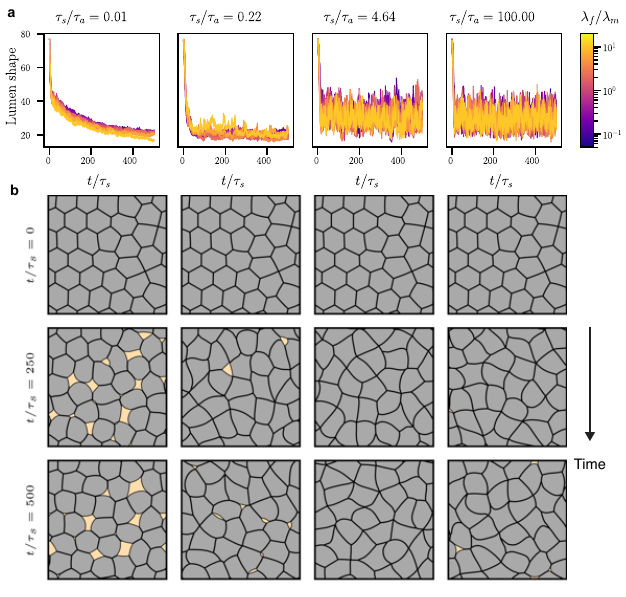}
    \caption{\textbf{Lumen shape analysis reveals microlumina do not coalesce on simulation timescales}. (a) Lumen Shape Parameter $\mathcal{A}_{\rm lumen} = p^2_{\rm lumen} / (4\pi a_{\rm lumen})$ for an interconnected lumen with perimeter $p_{\rm lumen}$ and area $a_{\rm lumen}$. $\mathcal{A}_{\rm lumen}$ has a minimum value of $1$, which indicates a purely circular lumen. Values of $\mathcal{A}_{\rm lumen} > 10$ indicate highly non-circular, extended lumina. Panels indicate the value of the timescale ratio $\tau_s/\tau_a$, as denoted by the panel titles, and curve color indicates the permeability ratio $\lambda_f/\lambda_m$, as denoted by the colorbar. (b) Example snapshots of simulations across the same values of $\tau_s/\tau_a$ shown in panel $a$ (columns) at different points in time (rows). }
    \label{fig:ExtDataFig_NoVoidCoalescence}
\end{figure*}

\pagebreak
\begin{figure*}[h!]
    \centering
    \includegraphics[width=\linewidth]{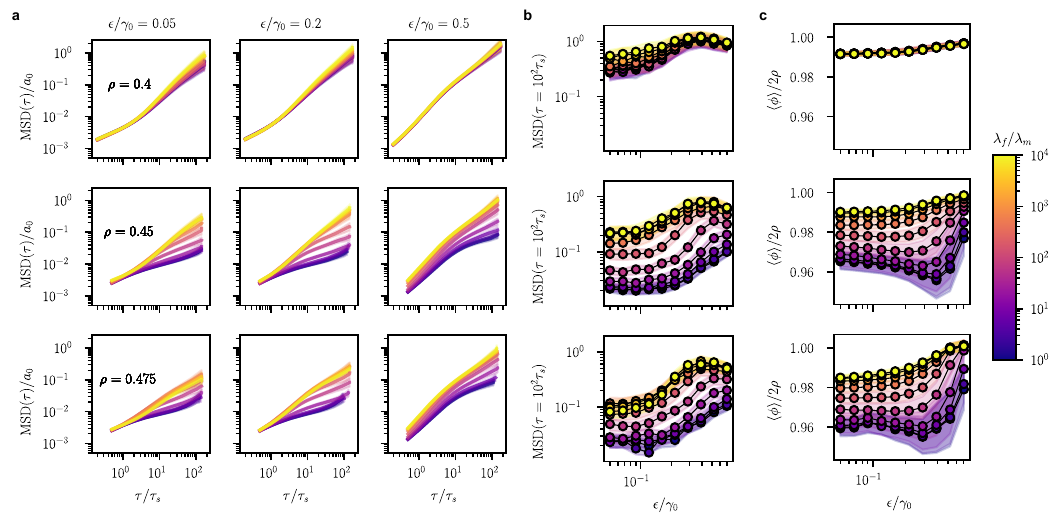}
    \caption{\textbf{Active tissues at lower density do not generate microlumina, but are sensitive to changes to permeability}. (a) MSD as a function of lag time $\tau$ scaled by the shape relaxation timescale $\tau_s$ of cell centers across different values of scaled number density $\rho$ (rows), adhesion ratio $\epsilon/\gamma_0$ (columns) and permeability ratio $\lambda_f/\lambda_m$ (see colorbar at right). (b) Same as panel (a), but with the MSD at a fixed lag time $\tau = 10^2 \tau_s$. Notice that simulations lower density $\rho=0.4$ are less sensitive to changes to permeability than simulations at higher densities, but changes to the permeability ratio still have an effect. (c) Same as (b), but now the steady state observed packing fraction $\langle \phi \rangle$ scaled by $2\phi$, which is the expected packing fraction for cells with areas near $2a_0$. Note that no discernible difference can be seen in $\langle \phi\rangle$ at $\rho=0.4$, indicating a lack of microlumina.}
    \label{fig:ExtDataFig_MSDComp}
\end{figure*}


\clearpage

\onecolumngrid

\setcounter{equation}{0}
\setcounter{figure}{0}
\setcounter{table}{0}
\setcounter{page}{1}
\setcounter{section}{0}
\makeatletter
\renewcommand{\theequation}{S\arabic{equation}}
\renewcommand{\thefigure}{S\arabic{figure}}
\setcounter{secnumdepth}{2} 
\renewcommand{\thesection}{S\arabic{section}}
\renewcommand{\thesubsection}{S\arabic{section}.\arabic{subsection}}
\makeatletter
\renewcommand{\p@subsection}{}
\makeatother

\begin{center}
{\LARGE Supplement for}\\
\vspace{0.4cm}
\textbf{\Large Flow, dynamics and active fracture in hydraulic multicellular systems}\\
\hspace{0.5cm}
\end{center}

\begin{center}

{\large
John D. Treado$^{1,2}$
Arthur Boutillon$^{1}$, 
Frank Jülicher$^{1,2,3}$, 
Otger Campàs$^{1,3,4}$
}

\vspace{0.3cm}

{\small
$^{1}$Cluster of Excellence Physics of Life, TU Dresden, 01062 Dresden, Germany\\
$^{2}$Max Planck Institute for the Physics of Complex Systems, 01187 Dresden, Germany\\
$^{3}$Center for Systems Biology Dresden, 01307 Dresden, Germany\\
$^{4}$Max Planck Institute of Molecular Cell Biology and Genetics, 01307 Dresden, Germany
}

\end{center}

\secttoc  

\section{Model Details}\label{suppsec:hydraulic_tess_model}

\begin{figure}[t]
    \centering
    \includegraphics[width=0.9\linewidth]{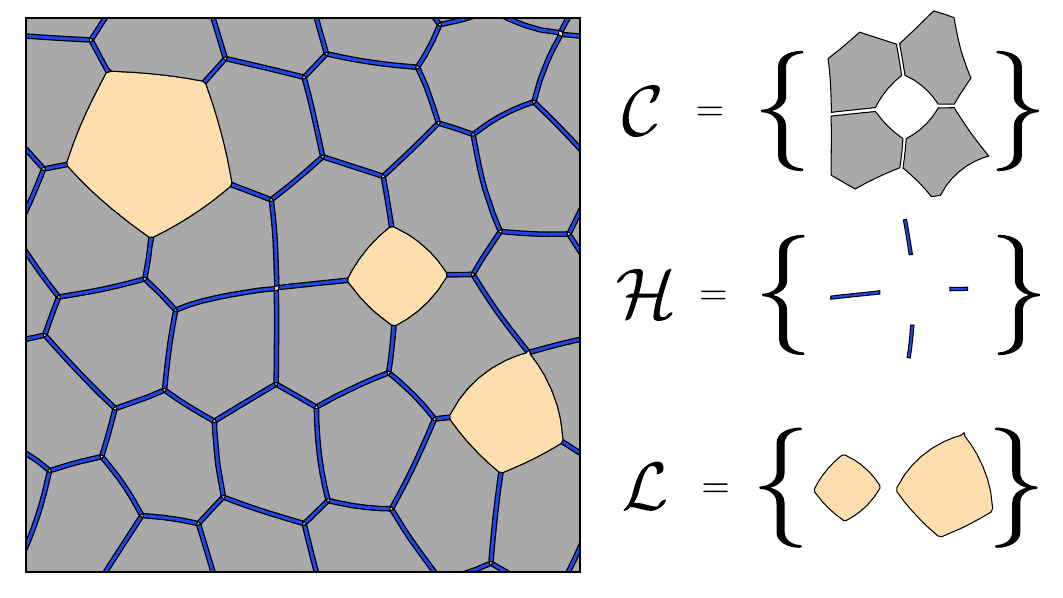}
    \caption{\textbf{Definition of partition sets in model}. From a particular configuration (left), the system is partitioned into sets of cells $\mathcal{C}$, channels $\mathcal{H}$, and lumina $\mathcal{L}$. }
    \label{suppfig:TessDef}
\end{figure}

In this section, we provide further details on our model of hydraulic partitions and interacting cells introduced in the Main Text. We consider a system of $N$ two dimensional cells
represented as deformable polygons. The system is also comprised of comprised of \emph{interstitial polygons} that exist in between the 
constituent cells. Interstitial polygons are obtained via a tessellation algorithm explained below in Section~\ref{suppsec:hydraulic_tess_model:tess_alg}.

As shown in Fig.~\ref{suppfig:TessDef}, let $\mathcal{C}$ indicate the set of all cells in the system, and let $\mathcal{I}$ indicate the set of all interstitial polygons subdivided into the \emph{channel set} $\mathcal{H}$ (blue regions in Fig.~\ref{suppfig:TessDef}) and the \emph{lumen set} $\mathcal{L}$ (beige regions in Fig.~\ref{suppfig:TessDef}). In all cases, we consider a boundary of fixed size, such that the areas $a$ of all polygons sum to the area of the boundary. For a square domain with side length 
$L$, total volume conservation dictates that
\begin{equation}
    \sum_{\mu \in \mathcal{C}} a_\mu + \sum_{q \in \mathcal{I}} a_k = L^2
\end{equation}
Here, $\mu \in \mathcal{C}$ denotes a cell $\mu$ taken from the total set $\mathcal{C}$, and $q \in \mathcal{I}$ denotes an interstitial polygon taken from the total set $\mathcal{I}$. We will use greek subscripts to denote cell indices, and latin subscripts to denote interstitial polygon indices.

\subsection{Mass Conservation}\label{suppsec:hydraulic_tess_model:mass_cons}
We begin with the mass conservation equation for a cell $\mu$. The total mass $M_\mu$ contained within the cell obeys the continuity equation
\begin{equation}
    \dot{M}_\mu = \sum_{q \in \mathcal{I}_\mu} J_{\mu q}.
\end{equation}
where we use a dot $\dot{}$ to signify total differentiation in time. Here, $J_{\mu q}$ is the mass flux between cell $\mu$ and interstitial polygon $q$, and we define $\mathcal{I}_\mu$ as the set of interstitial polygons (either channels or lumina) that are adjacent to cell $\mu$. We take the convention that $J_{\mu q} > 0$ means mass is entering the cell, and vice versa. By mass conservation, $J_{q \mu} = -J_{\mu q}$.
 
Note that there is no mass flux between cells, i.e. $J_{\mu \nu} = 0$ for $\nu \in \mathcal{C}$. However, mass can transfer from one interstitial polygon to another, i.e. $J_{q r} \neq 0$ for $q, r \in \mathcal{I}$. Therefore, the mass $M_q$ in interstitial polygon $q$ obeys the continuity equation
\begin{equation}
    \dot{M}_q = \sum_{\mu \in \mathcal{C}_q} J_{q\mu} + \sum_{r \in \mathcal{I}_q} J_{qr}.
\end{equation}
where $\mathcal{C}_q$ is the set of cells adjacent to interstitial polygon $q$, and $\mathcal{I}_q$ is the set of interstitial polygons adjacent to interstitial polygon $q$. By mass conservation, $J_{qr} = -J_{rq}$. 

We assume that the fluid that transits between interstitial polygons and cells is incompressible with constant mass density $\rho$. Then, the mass in a certain cell $\mu$ or interstitial polygon $q$ is related to the 
accompanying containing area $a_\mu$ or $a_q$ by
\begin{equation}
    M_\mu = \rho a_\mu, \quad M_q = \rho a_q.
\end{equation}
We can therefore rewrite the mass conservation equations as
\begin{subequations}
    \begin{align}
        \rho\dot{a}_\mu &= \sum_{q \in \mathcal{I}_\mu} J_{\mu q}, \\
        \rho\dot{a}_q &= \sum_{\mu \in \mathcal{C}_q} J_{q\mu} + \sum_{r \in \mathcal{I}_q} J_{qr}.  
    \end{align}
\end{subequations}
In the following, we will absorb the constant $\rho$ in the mass fluxes, and neglect writing it explicitly in future equations. 

\subsection{Constitutive Equations}\label{suppsec:hydraulic_tess_model:const_eqs}
We take the mass flux between cell $\mu$ and interstitial polygon $q$ to be controlled by two factors: hydrostatic pressure differences and osmotic pressure differences. We consider each cell $\mu$ and interstitial polygon $q$ to have hydrostatic pressures $P_\mu$ and $P_q$, respectively, and osmotic pressures $\Pi_\mu$ and $\Pi_q$, respectively. We assume that hydrostatic pressure $P$ and osmotic pressure $\Pi$ are homogeneous within each fluid partition. 
We write the mass flux between cell $\mu$ and interstitial polygon $q$ as
\begin{equation}\label{suppeq:cell_flux}
    J_{\mu q} =  \ell_{\mu q}\qty(\lambda_m \qty[\qty(P_q - P_\mu) - \qty(\Pi_q - \Pi_\mu)] + j_{\mu q}).
\end{equation}
Here, $\lambda_m$ is a constant membrane permeability to flow per unit length, and $\ell_{\mu q}$ is the length of the interface between cell $\mu$ and interstitial polygon $q$. The flux density $j_{\mu q}$ represents active fluid transport; in this paper, we will set this term $j_{\mu q} = 0$, but note that active transport through ion pumps can play a role in tissue hydraulics.

In our model, we do not explicitly track osmolyte concentration differences in difference partitions; we take the osmotic pressure of each cell $\mu$ to have the phenomenological form
\begin{equation}
    \Pi_\mu = \frac{\kappa}{a_\mu - a_{0\mu}}
\end{equation}
where $\kappa$ is an energy that sets the scale of the internal cell osmotic pressure, and $a_{0\mu}$ is the area taken up by solid material within the cell, which sets a lower-bound for the cell area. This model for cell osmotic pressure is taken from experimental observations of cell response to osmotic shocks~\cite{Guo.Weitz.2017}, and assumes that osmotic pressure in a cell is controlled by trapped, non-interacting osmolytes that can only diffuse in a liquid area $a_\mu - a_{0\mu}$ within a cell. In general, we fix the value of $a_{0\mu} = a_0$, but can introduce polydispersity by drawing each $a_{0\mu}$ as a random number from a distribution, as shown in Fig.~\ref{fig:Fig3} in the Main Text. We assume the osmotic pressure of interstitial space takes on a constant value $\Pi_q = \Pi_0$ for any $q \in\mathcal{I}$. Because osmotic pressure is constant in all interstitial polygons, the mass flux between connected interstitial polygons is given solely by hydrostatic pressure differences. Therefore
\begin{equation}\label{suppeq:fluid_flux}
    J_{qr} = \lambda_f\ell_{qr}\qty(P_r - P_q).
\end{equation}
Here, $\ell_{qr}$ is the length of the interface between interstitial polygons $q$ and $r$, and $\lambda_f$ is the permeability of the interface between interstitial polygons. 
This parameter can be seen as an inverse measure of the resistance to flow, which is controlled by the viscosity of the fluid and porosity of the interstitial space.

\subsection{Momentum Conservation}\label{suppsec:hydraulic_tess_model:mom_cons}
We assume that the vertices of each polygon undergo forces generated by an effective potential energy $E$, which we take to be 
\begin{equation}
    E = E_{\rm int} + \sum_{\mu \in \mathcal{C}} (U_\mu - P_\mu a_\mu) - \sum_{q \in \mathcal{I}} P_q a_q.
\end{equation}
Here, $U_\mu$ is the effective internal surface energy that generates cortical forces on cell $\mu$, and $E_{\rm int}$ is the effective energy associated with polygonal surface-surface interactions. For each cell $\mu$, we assume that there is a cortical tension $\gamma_i$ that can vary, in principle, over each edge $i$ on the polygonal surface of cell $\mu$. Let $E_\mu$ be the set of edges that define cell $\mu$. The effective internal surface energy of cell $\mu$ is defined as
\begin{equation}
    U_\mu = \sum_{i \in E_\mu} \gamma_i l_i
\end{equation}
where $l_i$ is the length of polygonal edge $i$. The interaction energy $E_{\rm int}$ is defined and described in more detail in Sec.~\ref{suppsec:adhesion}. 

We assume the vertices $i$ of each polygon (either cell or interstitial) experience a frictional force $\vb*{F}^\zeta_i = -\zeta \vb*{v}_i$ given a vertex velocity $\vb*{v}_i$. Given that the total force on vertex $i$ is
\begin{equation}
    \vb*{F}_i = -\dv{U}{\vb*{r}_i} - \zeta \vb*{v}_i,
\end{equation}
momentum conservation in the overdamped limit gives
\begin{equation}\label{suppeq:mom_cons_all_forces}
    \zeta \dot{\vb*{r}_i} = \vb*{\gamma}_i + \vb*{\epsilon}_i + \vb*{P}_i 
\end{equation}
where we assume vertex $i$ is a part of cell $\mu$, $\mathcal{I}_i$ is the set of interstitial polygons adjacent to vertex $i$, and we have defined 
\begin{subequations}
    \begin{align}
        \vb*{\gamma}_i &= -\dv{U_\mu}{\vb*{r}_i} \\
        \vb*{\epsilon}_i &= -\dv{E_{\rm int}}{\vb*{r}_i}\\
        \vb*{P}_i &= P_{\mu}\dv{a_\mu}{\vb*{r}_i} + \sum_{q \in \mathcal{I}_i} P_q \dv{a_q}{\vb*{r}_i}
    \end{align}
\end{subequations}
as the forces on vertex $i$ due to cortical tension, adhesion, and hydrostatic pressure, respectively. Note that the chain rule gives
\begin{equation}
    \pmb{\gamma}_i = \gamma_i \hat{\pmb{l}}_i - \gamma_{i-1}\hat{\pmb{l}}_{i-1}.
\end{equation}
Thus, the contributions to this force shown in Fig.~\ref{fig:Fig1}d are defined as $\pmb{\gamma}_{i,i} = \gamma_i \hat{\pmb{l}}_i$ and $\pmb{\gamma}_{i,i-1} =  - \gamma_{i-1}\hat{\pmb{l}}_{i-1}$.

For the hydrostatic pressure force $\vb*{P}_i$, we provide a geometric interpretation of the quantity $da_\mu /d\vb*{r}_i$. Using the ``shoestring" formula for a closed $n$-gon area
\begin{equation}
    a_\mu = \frac{1}{2}\sum_{i=1}^n x_i y_{i+1}-x_{i+1}y_i
\end{equation}
the derivative $da_\mu /d\vb*{r}_i$ gives
\begin{equation}
    \dv{a_q}{\vb*{r}_i} = \qty(\frac{y_{i+1}-y_{i-1}}{2}, \frac{x_{i-1}-x_{i+1}}{2}).
\end{equation}
If we define $\vb*{s}_i$ as the vector connecting vertex $i-1$ and $i+1$, i.e.
\begin{equation}
    \vb*{s}_i = \qty(x_{i+1}-x_{i-1}, y_{i+1}-y_{i-1}),
\end{equation}
then the vector above can be seen as 
\begin{equation}
    \dv{a_\mu}{\vb*{r}_i} = \frac{\vb*{s}_i^\perp}{2}
\end{equation}
where we define the \emph{orthogonal complement} $\perp$ to a particular vector as 
\begin{equation}
    \vb*{v} = (v_x, v_y) \quad ; \quad \vb*{v}^\perp = (v_y, -v_x).
\end{equation}
We then define
\begin{equation}\label{eq:outward_ortho}
    \vb*{o}_{i\mu} \equiv \dv{a_\mu}{\vb*{r}_i}.
\end{equation}
Because $\perp$ orthogonalizes a vector, $\vb*{o}_{i\mu}$ can be seen as a vector with length $|(l_{i}+l_{i-1})/2|$ but pointing normal to the cell surface at vertex $i$. The vectors $\vb*{o}_{i\mu}$ are thus defined relative to the surface of a particular partition $\mu$. Consider a vertex $i$ connected to only two partitions, a cell $\mu$ and the fluid $q$. For cell $\mu$, the order of vertices $i-1$, $i$ and $i+1$ is reversed for that order on the partition $q$. Tn this case we therefore have
\begin{equation}
    \vb*{o}_{i\mu} = -\vb*{o}_{iq}
\end{equation}
and the net pressure force $\vb*{P}_i$ on vertex $i$ is given by
\begin{equation}
    \vb*{P}_i = (P_\mu - P_q)\vb*{o}_{i\mu},
\end{equation}
i.e. a force that points normal to the surface of cell $\mu$ at vertex $i$ and is controlled by the pressure difference $P_\mu - P_q$. 

\subsection{Constraint Satisfaction}\label{suppsec:hydraulic_tess_model:constrain_sat}
In order to satisfy both momentum and mass conservation simultaneously, we determine the hydrostatic pressures $P_\mu$ and $P_q$ in each cell and interstitial polygon as Lagrange multipliers. First, we note that a time derivative of any polygon, be it a cell $\mu$ or interstitial polygon $q$, can be expressed by the chain rule as
\begin{subequations}\label{eq:constraint}
    \begin{align}
        \dot{a}_\mu &= \sum_{i \in V_\mu} \dot{\vb*{r}}_i \cdot \dv{a_\mu}{\vb*{r}_i} = \sum_{i \in V_\mu} \dot{\vb*{r}}_i \cdot \vb*{o}_{i\mu}\\
        \dot{a}_q &= \sum_{i \in V_q} \dot{\vb*{r}}_i \cdot \dv{a_q}{\vb*{r}_i} = \sum_{i \in V_\mu} \dot{\vb*{r}}_i \cdot \vb*{o}_{iq}
    \end{align}
\end{subequations}
where we have defined $V_\mu$ and $V_q$ to be the set of vertices in cell $\mu$ and interstitial polygon $q$, respectively. We have also used the outward orthogonal vectors $\vb*{o}_{i\mu}$ defined in Eq.~\eqref{eq:outward_ortho} above. Using Eq.~\eqref{suppeq:mom_cons_all_forces}, we can write the rate of change of the area of a given polygon (e.g. cell $\mu$) as 
\begin{equation}
    \dot{a}_\mu = \zeta^{-1}\qty(\Phi_\mu + P_\mu O_\mu + \sum_{q \in \mathcal{I}_\mu} P_q O_{\mu q})
\end{equation}
where we have defined the quantities
\begin{subequations}\label{suppeq:quant_def}
    \begin{align}
        \Phi_\mu &= \sum_{i \in V_\mu} \qty(\vb*{\gamma}_i + \vb*{\epsilon}_i) \cdot \vb*{o}_{i\mu} \\
        O_\mu &= \sum_{i \in V_\mu} \vb*{o}_{i\mu} \cdot \vb*{o}_{i\mu} \\
        O_{\mu q} &= \sum_{i \in \ V_\mu \cap V_q} \vb*{o}_{i\mu} \cdot \vb*{o}_{iq}
    \end{align}
\end{subequations}
The first quantity, $\Phi_\mu$ is the work done by applying the non-pressure forces in the direction $da_\mu/d\vb*{r}_i$, which is normal to the surface of $\mu$ at the vertex $i$. The second two terms have units of area, and note that $O_{\mu q}$ is defined only at the interface between a given cell $\mu$ and interstitial fluid partition $q$. 

Using the constitutive relations in Eqs.~\eqref{suppeq:cell_flux} and~\eqref{suppeq:fluid_flux}, we arrive at the linear system 
\begin{equation}\label{suppeq:lin_sys}
    A \vb{P} = \vb{\Omega}
\end{equation}
where $\vb{P}$ is a vector of pressures in each fluid partition, $\vb{\Omega}$ is a vector with units of energy assigned to each partition, and $A$ is a matrix with units of area with elements
\begin{equation}
    A_{kl}=\begin{cases}
        O_k + \zeta \lambda_m p_k & \text{ if } k = l, k \in \mathcal{C} \\
        O_{k l} - \zeta\lambda_m \ell_{k l} & \text{ if } k \neq l, k \in \mathcal{C}, l\in\mathcal{I}\\
        O_k + \zeta\lambda_f\sum_{r \in \mathcal{I}_k}\ell_{kr} + \zeta\lambda_m\sum_{\mu \in \mathcal{C}_k}\ell_{k\mu} & \text{ if } k = l, k \in \mathcal{I}\\
        O_{k l} - \zeta\lambda_m \ell_{k l} & \text{ if } k \neq l, k \in \mathcal{I}, l \in \mathcal{C} \\
        O_{k l} - \zeta\lambda_f \ell_{k l}&\text{ if } k \neq l, k \in \mathcal{I}, l \in \mathcal{I}
    \end{cases}
\end{equation}
where $\ell_{kl}$ is the interfacial length between partition $k$ and partition $l$. The vector $\vb{\Omega}$ is defined as
\begin{equation}
    \Omega_k = \begin{cases}
        \zeta  p_k (j_{k} -\lambda_m\Delta \Pi_k)-\Phi_k & \text{ if } k\in \mathcal{C} \\
        \zeta\left[\sum_{\mu \in \mathcal{C}_k}\ell_{\mu k}(\lambda_m \Delta \Pi_\mu - j_{\mu})\right] - \Phi_k & \text{ if } k \in \mathcal{I}.
    \end{cases}
\end{equation}
where the work $\Phi_k$ is defined above in Eqs.~\eqref{suppeq:quant_def}. 

\subsection{Tessellation Algorithm}\label{suppsec:hydraulic_tess_model:tess_alg}
\begin{figure}
    \centering
    \includegraphics[width=0.95\linewidth]{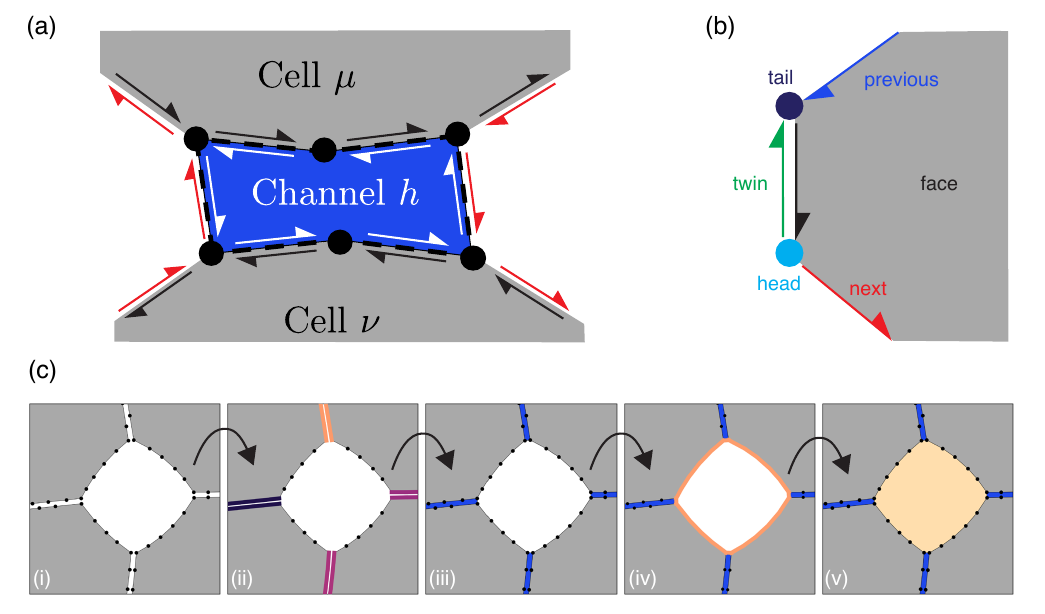}
    \caption{\textbf{Half-edge data structure and Interstitial Tessellation algorithm}. (a) Half-edges drawn at a cell-cell interface with one channel. Half-edges can either be interior to cells (black), interior to channels (white), or exterior to both (red). (b) Every half-edge (black) points to the previous edge in the cycle (blue), the next edge in the cycle (red), a twin edge that points in the opposite direction (green), the face of which the half-edge is a part (gray), and the vertex at the tail end of the half-edge (purple). Note the half-edge need not point at the vertex at its head, it need only ask the tail vertex of its twin. (c) Schematic of interstitial tessellation algorithm employed in this work. (i) From an initially empty interstitial space defined by the negative space of a packing of cells, (ii) clusters of interacting edges are determined and unified to make (iii) distinct channels. (iv) Lumina are then identified by the space surrounded by non-unified cell edges and the exterior boundaries of channels, making (v) distinct spaces for lumina and channels. }
    \label{suppfig:TessAlgorithm}
\end{figure}

In order to simulate fluid flux between adjacent partitions, partitions and their neighbors need to be identified. To represent partitions as connected polygons in simulations, we use use the half-edge data structure commonly employed in surface meshes for computational graphics~\cite{Kettner.Kettner.1999}. The general idea of the data structure is shown in Fig.~\ref{suppfig:TessAlgorithm}a; polygon edges are decomposed into directional ``half-edges" that connect adjacent vertices and can be uniquely assigned to an individual vertex and an individual polygonal face (see Fig.~\ref{suppfig:TessAlgorithm}b). For a more detailed introduction, see Section 3.2 in Ref.~\cite{Kettner.Kettner.1999}. 

At the beginning of each time step, we initialize a set of $2N_V$ half-edges associated with the $N_V$ total vertices in the system. Of this set, half are bound to the interiors of cells (see black half-edges in Fig.~\ref{suppfig:TessAlgorithm}a) and wind themselves counter-clockwise, while the rest are bound on the exterior of cells and wind themselves clockwise (see red half-edges in Fig.~\ref{suppfig:TessAlgorithm}a). The winding order is important, as the creation of additional polygonal faces in the interior of a set of cells (e.g. a channel) will have interior half-edges that also wind counter-clockwise and exterior edges that wind clockwise (see white half-edges in Fig.~\ref{suppfig:TessAlgorithm}a). 

Once the initial set of half edges are initialized at the beginning of a time step, forces due to shape ($U_\mu$) and interaction ($U_{\rm int}$) potentials are computed for a given configuration. If a vertex $i$ interacts with an edge $j$, and if the midpoint of the edge associated to vertex $i$ projects onto edge $j$, or vice versa, then the exterior half-edges bound to those edges are added to the same cluster via the Newman-Ziff labeling algorithm~\cite{Newman.Ziff.2001}. We proceed until all interacting edge clusters are labeled (see Fig.~\ref{suppfig:TessAlgorithm}c-ii). 

Once cluster labels are determined, the set of exterior half-edges that are part of the same cluster are assigned to a new face, which is an interfacial channel (see Fig.~\ref{suppfig:TessAlgorithm}c-iii). This requires two new edges (four half-edges) to be created at the boundary of the interfacial region, and rewiring of the relevant half-edges to keep the network of half-edges consistent. Once channels are created, lumina are identified as the faces created by the sets of external half-edges \emph{not} belonging to any existent cluster, and connected by the exterior half-edges of the newly created channels. We traverse all closed loops in the system that are formed in this way, and associated each half-edge that is a part of each closed loop to a new polygonal face, i.e. the lumen. 

All created channels necessarily contain half-edges that wind counter-clockwise, but connected interfaces can have winding of either handedness. For example, in Fig.~\ref{suppfig:TessAlgorithm}a, the red half-edges wind \emph{clockwise} and would encircle the exteriors of two cells that form the adhesive interface, thus creating not an interior lumen but an exterior void. On the other hand, connected free edges formed by the boundary of cell and channel interfaces that are interior to the tissue would have counter-clockwise handedness. We therefore use the handedness of connected interfaces to determine whether cell and channel boundaries create interior lumina or exterior voids.

\newpage 
\section{Interaction Potential}\label{suppsec:adhesion}

\begin{figure}[!h]
    \centering
    \includegraphics[width=\linewidth]{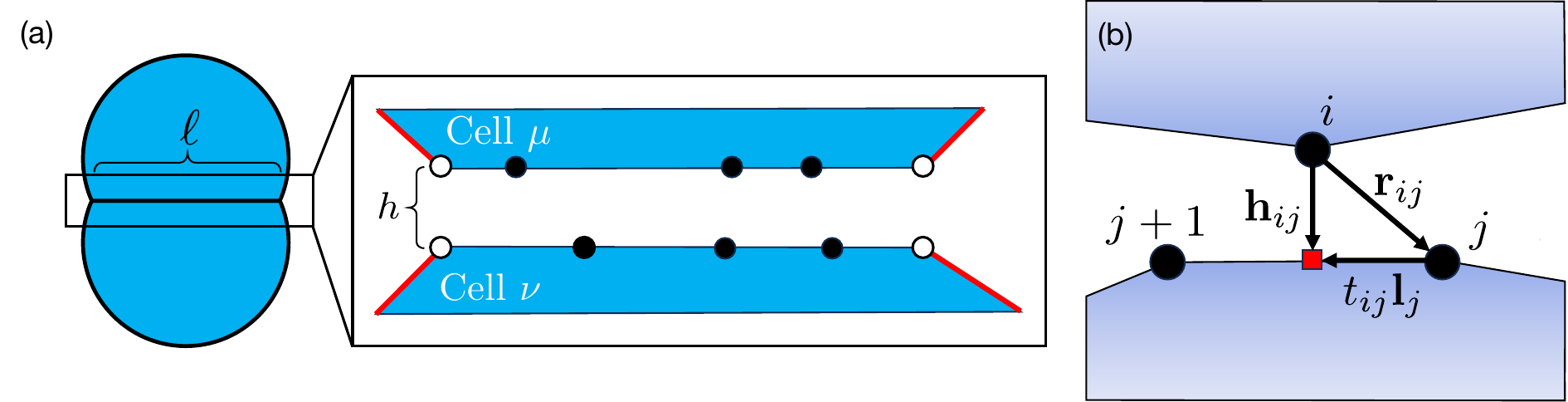}
    \caption{(a) Heuristic scenario for the definition of the interaction potential. Two cells $\mu$ and $\nu$ form an adhesive interface of length $\ell$ (left) if the microscopic arrangement of vertices and edges (right) are locally flat, and only parallel edges (black) are counted as part of the interface while non-parallel edges (red) are ignored. (b) Schematic defining the geometry of the projection distance $h_{ij}$ and projection coordinate $t_{ij}$ for a vertex $i$ interacting with the edge defined by the distance between vertices $j$ and $j+1$. Vertices are indicated by black circles, and a red square indicates the point where the vertex $i$ projects onto the edge between vertices $j$ and $j+1$. }
    \label{suppfig:int_potential_def}
\end{figure}
In this section, we detail the microscopic interaction potential chosen to model adhesive cell interfaces. This problem is motivated by the heuristic scenario depicted in Figure~\ref{suppfig:int_potential_def}a, where two cells $\mu$ and $\nu$ form an interface of length $\ell$. The cortical tension $\gamma$ tends to minimize the cell perimeter unless there is an additional interfacial energy $U_{\rm int}$ that seeks to drive the interface to grow and spread on another cell. For an interface of length $\ell$, the interfacial energy should have the form
\begin{equation}\label{eq:macro_adh}
    E_{\rm int} = -2\ell W
\end{equation}
where $W > 0$ is the adhesion. By tuning the ratio $W/\gamma$, one can tune the equilibrium interfacial size $\ell^*$. 

We therefore construct a microscopic interaction potential $E_{\rm int}$ that is (a) a function of distances between polygonal vertices, such that it can be evaluated for any arbitrary polygon configuration, but also (b) reproduces the macroscopic heuristic of an interfacial energy, i.e. $E_{\rm int}=-2\ell W$, when two cells meet at a flat interface of length $\ell$ (Fig.~\ref{suppfig:int_potential_def}a). In this section, we will show that a microscopic interaction potential that satisfies these requirements can be written in the form
\begin{equation}\label{suppeq:int_pot_def}
    E_{\rm int} = V_{\rm vv} + \sum_{i \in V} \sum_{j \in E} \bar{l}_{ij}   w(t_{ij}) \varphi(h_{ij}).
\end{equation}
where $V$ ($E$) is the set of all vertices (edges) throughout the system. The first term in Eq.~\eqref{suppeq:int_pot_def}, $V_{\rm vv}$, is a specific vertex-vertex interaction potential that is only required for potential continuity, and is described in the section~\ref{suppsec:adhesion:vert-vert}. The second term is a vertex-edge interaction potential and is comprised of three ingredients: $\bar{l}_{ij}$, which is the \emph{effective local surface element} for the interaction between vertex $i$ and edge $j$, $\varphi(h_{ij})$, which is an interaction potential energy \emph{density} that is a function of the \emph{projection distance} $h_{ij}$ of vertex $i$ onto edge $j$, and an activation function $w$ that controls the activation or inactivation of the potential depending on the \emph{projection coordinate} $t_{ij}$. These quantities are defined and described in detail in the following text. 

\subsection{Effective Local Surface Element $\bar{l}_{ij}$}\label{suppsec:adhesion:surface-element}
In order to define a microscopic potential that reproduces a macroscopic adhesion energy, i.e. $E_{\rm int} = -2\ell W$, we must incorporate local surface information near interacting vertices and edges. However, those surface elements must only be a part of the surface; edges near the boundaries of interfaces must only be considered if they indeed lie on the interface, and must be ignored if not. 

To this end, we define the effective local surface element $\bar{l}_{ij}$ of vertex $i$ interacting with edge $j$  as 
\begin{equation}
    \bar{l}_{ij} = \frac{l_i \Xi_{ij} + l_{i-1}\Xi_{i-1, j}}{2}.
\end{equation}
Here, $l_i$ is the length of the edge connecting vertex $i$ and vertex $i+1$. The matrix $\Xi_{ij}$ is a boolean matrix that determines if the edge connecting vertices $i$ and $i+1$ are a member of the same channel element (as described in Sec.~\ref{suppsec:hydraulic_tess_model}) as edge $j$. That is, we define
\begin{equation}
    \Xi_{ij} = \begin{cases}
        1 & \text{ if } \exists \  h \in \mathcal{H} \   : \  i, j \in E_h \\
        0 & \text{ otherwise}.
    \end{cases}
\end{equation}
With this definition, edges that contribute to the channel at the interface between to cells lend their edge length to the potential, whereas edges that ``hang" off of the side (as shown in red in Fig.~\ref{suppfig:int_potential_def}a) are not counted. Note then that we first identify the channel identities of each edge according to the interstitial tesselation algorithm described in Sec.~\ref{suppsec:hydraulic_tess_model:tess_alg}, then can only compute the force due to the adhesive potential. 

\subsection{Projection Distance $h_{ij}$ and Projection Coordinate $t_{ij}$}\label{suppsec:adhesion:proj-dist-and-coord}
The projection distance $h_{ij}$ from vertex $i$ to edge $j$ (that is, the edge between vertices $j$ and $j+1$) is shown schematically Figure~\ref{suppfig:int_potential_def}b. We define this distance as the magnitude of the vector $\vb*{h}_{ij}$, which is defined 
\begin{equation}
    \vb*{h}_{ij} = \vb*{r}_{ij} + t_{ij}\vb*{l}_j
\end{equation}
where $\vb*{r}_{ij} = \vb*{r}_j - \vb*{r}_i$ is the distance between the positions of vertices $j$ and $i$, and $\vb*{l}_j = \vb*{r}_{j+1} - \vb*{r}_j$ is a vector that points from vertex $j$ to $j+1$. The quantity $t_{ij}$, introduced above, is a dimensionless coordinate that describes how far along the projection point of vertex $i$ onto edge $j$ points along edge $j$. This is defined as
\begin{equation}
    t_{ij} = -\frac{\vb*{r}_{ij} \cdot \vb*{l}_j}{\vb*{l}_j \cdot \vb*{l}_j}.
\end{equation}
With this definition, $\vb*{h}_{ij} \cdot \vb*{l}_j = 0$, i.e. the projection vector $\vb*{h}_{ij}$ is always perpendicular to the edge vector $\vb*{l}_j$. 

If $t_{ij}$ lies within the interval $[0, 1]$, then the vertex $i$ has a projection that lies \emph{on} the edge $j$, i.e. the point of projection $(1-t_{ij})\vb*{r}_j + t_{ij}\vb*{r}_{j+1}$ lies between the coordinate $\vb*{r}_j$ and $\vb*{r}_{j+1}$. Otherwise, the vertex $i$ projects somewhere onto the line formed by $\vb*{l}_j$, but not between the vertices $j$ and $j+1$. Therefore, we define the activation function $w(t)$ as 
\begin{equation}
    w(t) = \begin{cases}
        1 \text{ if } t \in [0, 1] \\
        0 \text{ otherwise}.
    \end{cases}
\end{equation}
With this definition, the potential only is non-zero if a vertex $i$ projects onto the edge $j$. 

\subsection{Validation of the microscopic potential}\label{suppsec:adhesion:validation}
We now check whether the microscopic interaction potential introduced in Eq.~\eqref{suppeq:int_pot_def} satisfies the requirement $E_{\rm int}=-2\ell W$ for a flat interface. In this case, we assume all vertices are separated by a distance $h$ by all edges, and that, for each vertex $i$, there exists some edge $j$ such that $t_{ij}\in [0, 1]$. If the two cells at the interface are cells $\mu$ and $\nu$ then, according to Eq.~\eqref{suppeq:int_pot_def}, the interaction potential energy is
\begin{equation}\label{eq:pot_deriv_1}
    E_{\rm int} = V_{\rm vv} + \varphi(h)\qty[\sum_{i \in V_\mu} \sum_{j \in E_\nu} \bar{l}_{ij} w(t_{ij})  + \sum_{k \in V_\nu} \sum_{m\in E_\mu} \bar{l}_{km} w(t_{km}) ]
\end{equation}
where $V_\mu$ and $E_\mu$ are the sets of vertices and edges on cell $\mu$, respectively. If for some vertex $i$ $t_{ij} \in [0, 1]$ with at most one edge $j$, then each sum above in Eq.~\eqref{eq:pot_deriv_1} reduces to a sum over each vertex on each interface. If we further consider that $\bar{l}_{ij}$ only includes contributions from edges that are a member of the interfacial channel, then our sum reduces to
\begin{equation}
    E_{\rm int} = V_{\rm vv} + 2\ell \varphi(h)
\end{equation}
where $\ell$ is the total length of the interface. As we will describe below, the vertex-vertex interaction potential $V_{\rm vv}$ will only be necessary at curved regions a polygonal surface. Therefore, for a flat interface, the microscopic interaction potential energy $E_{\rm int}$ takes the form
\begin{equation}
    E_{\rm int} = 2 \ell \varphi(h).
\end{equation}
Thus the macroscopic adhesion $W$ can be identified as 
\begin{equation}
    W = -\varphi(h).
\end{equation}
This correspondence can be seen in Fig.~\ref{fig:Fig2}b of the Main Text, where the contact angle $\theta$ formed at the adhesive interface between two simulated cells follows the Law of Young-Dupré, which takes the form
\begin{equation}
    \cos \theta = \frac{\gamma - W}{\gamma}
\end{equation}
if we replace $-\varphi(h)$ by $W$. If our potential energy density has a minimum at some distance $\sigma_0$ such that $\varphi(\sigma_0)=-\epsilon<0$, then we can identify
\begin{equation}
    W = -\epsilon
\end{equation}
for equilibrium interfaces with a channel height determined by the minimum of the potential energy density.

\subsection{Force Decomposition}\label{suppsec:adhesion:force_decomp}
In the main text, we note that a force due to our adhesion potential has two contributions: a force $\vb*{\epsilon}^\perp$ that acts normal to a cell-cell interface and pulls one cell surface towards another, and a force $\vb*{\epsilon}^\parallel$ that acts parallel to the interface and causes that interface to increase in length. We now show how these forces are defined. Given the definition of the microscopic interaction potential $E_{\rm int}$ in Eq.~\eqref{suppeq:int_pot_def}, and ignoring the vertex-vertex interaction, the force on a vertex $k$, $\vb*{\epsilon}_k-\partial E/\partial \vb*{r}_i$, is given by
\begin{equation}\label{suppeq:potential_breakdown}
    \vb*{\epsilon}_k = -\sum_{i \in V} \sum_{j \in E} w(t_{ij}) \qty[\bar{l}_{ij}\pdv{\varphi}{\vb*{r}_k} + \varphi(h_{ij}) \pdv{\bar{l}_{ij}}{\vb*{r}_k}].
\end{equation}
Note that by the chain rule, we have
\begin{equation}
    \pdv{\varphi}{\vb*{r}_k} = \pdv{\varphi}{h_{ij}} \pdv{h_{ij}}{\vb{r}_k} = \pdv{\varphi}{h_{ij}} \vu*{h}_{ij}\qty[(1 - t_{ij})\delta_{jk} + t_{ij}\delta_{j+1,k}-\delta_{ik}].
\end{equation}
where $\delta_{ij}$ is the Kronecker-$\delta$ function
\begin{equation}
    \delta_{ij} = \begin{cases}
        1 &\text{ if } i = j\\
        0 &\text{ otherwise}
    \end{cases}
\end{equation}
and $\vu*{h}_{ij}$ is the unit vector that points in the $\vb*{h}_{ij}$ direction. The key here is that this derivative is a vector that points normal to the cell-cell interface in the direction of $\vu*{h}_{ij}$. Therefore we can define
\begin{equation}
    \vu*{\epsilon}_k^\perp = -\sum_{i \in V} \sum_{j \in E} w(t_{ij})\qty[\pdv{\varphi}{h_{ij}} \qty((1 - t_{ij})\delta_{jk} + t_{ij}\delta_{j+1,k}-\delta_{ik})]\vu*{h}_{ij}.
\end{equation}
We refrain from simplifying the sum, as it sheds no particular light on the structure of the force; the key point is that this contribution to the force is strictly normal to cell-cell interfaces because they all lie along the directions of $\vu*{h}_{ij}$. 

\begin{figure}[!t]
    \centering
    \includegraphics[width=1\linewidth]{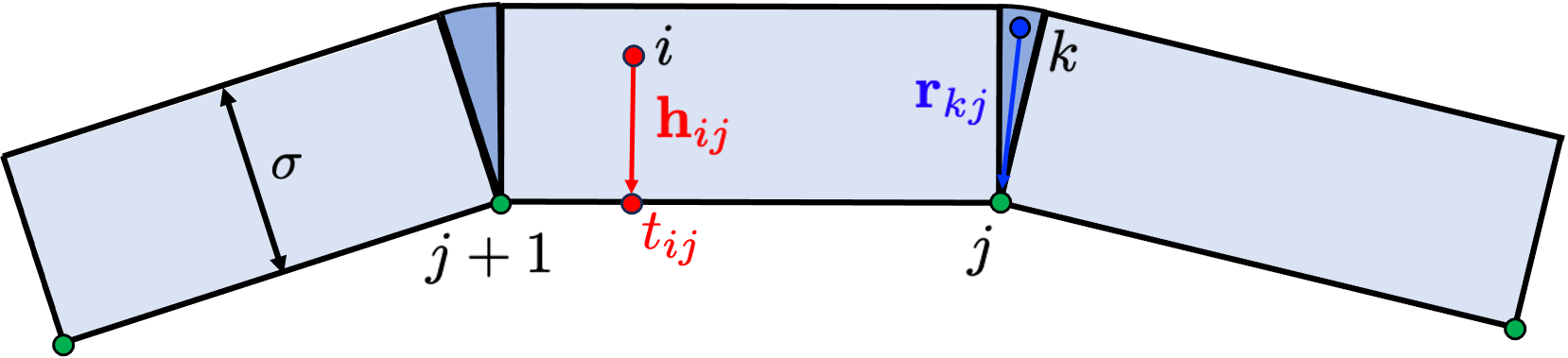}
    \caption{Schematic of a situation where an additional potential between vertices is required. The vertex $i$ has a projection $t_{ij} \in [0, 1]$ onto the edge $j$, and if it is within the interaction distance $\sigma$ of the potential energy density $\varphi$, it will interact with the edge. However, the vertex $k$ is within the distance $\sigma$ to the vertex $j$, yet $t_{ij} < 0$ and $t_{ij} > 1$.}
    \label{suppfig:vv_pot_required}
\end{figure}

The other term in Eq.~\eqref{suppeq:potential_breakdown} is given by
\begin{equation}
    \pdv{\bar{l}_{ij}}{\vb*{r}_k} = \frac{1}{2}\qty(\Xi_{ij}\pdv{l_i}{\vb*{r}_k} + \Xi_{i-1,j}\pdv{l_{i-1}}{\vb*{r}_k}) 
\end{equation}
which can be written
\begin{equation}
    \pdv{\bar{l}_{ij}}{\vb*{r}_k} = \frac{1}{2}\qty[\Xi_{ij}\qty(\delta_{i+1,k} - \delta_{ik})\vu*{l}_i + \Xi_{i-1,j}\qty(\delta_{ik} - \delta_{i-1,k})\vu*{l}_{i-1}].
\end{equation}
Thus, this term is only perpendicular to unit vectors $\vu{l}_i$ pointing along edges $i$. Since the matrix $\Xi_{ij}$ is only non-zero for edges that participate in the channel, this derivative points necessarily along the cell-cell interface. Thus, we can identify 
\begin{equation}
    \vu*{\epsilon}_k^\parallel = -\frac{1}{2}\sum_{i \in V} \sum_{j \in E} w(t_{ij})\qty[\varphi(h_{ij})\qty[\Xi_{ij}\qty(\delta_{i+1,k} - \delta_{ik})\vu*{l}_i + \Xi_{i-1,j}\qty(\delta_{ik} - \delta_{i-1,k})\vu*{l}_{i-1}]].
\end{equation}

\subsection{Vertex-vertex potential}\label{suppsec:adhesion:vert-vert}

\begin{figure}[t]
    \centering
    \includegraphics[width=0.7\linewidth]{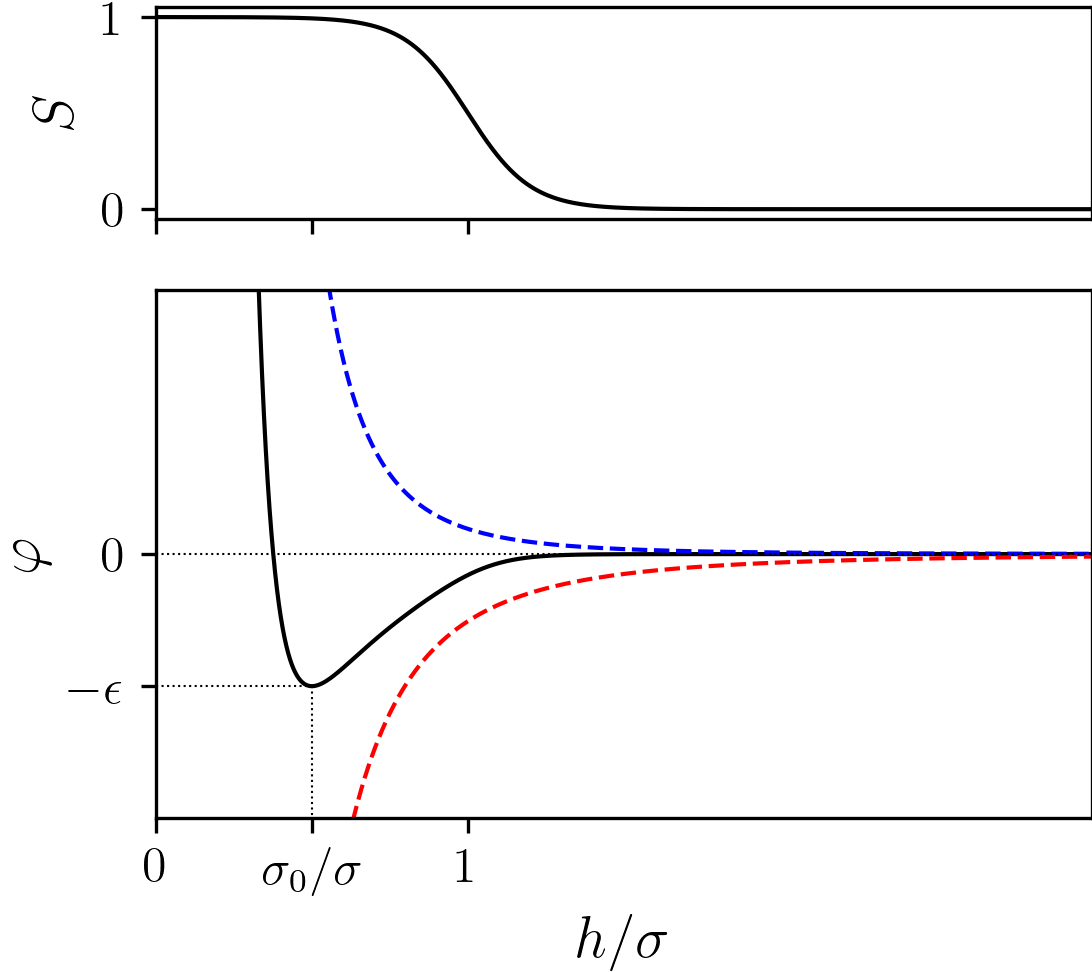}
    \caption{\textbf{Potential energy density $\varphi(h)$}. The pair potential density $\varphi$ chosen in this work is defined in Eq.~\eqref{suppeq:giammona_campas_pot} with $\alpha = 4$ and $\beta = 3$. The support function $S$ is shown in the top plot with steepness $a = 10^{-1}\sigma$, and the total potential density $\varphi$ as well as the attractive and repulsive parts of $B(h)$ are shown in the bottom plot in black, red and blue, respectively. Here, $\sigma_0=\sigma/2$, which is the value used throughout this work. }
    \label{suppfig:pot_en_dens}
\end{figure}

As shown in Fig.~\ref{suppfig:vv_pot_required}, there are occasions where a vertex is within a distance from the surface a cell, yet if only vertex-edge interactions were considered, no interaction would take place. This could be dangerous for a simulation and lead to cellular interpenetration, which would eventually cause diverging energies and a breakdown of the interstitial tesselation algorithm. 

Therefore, we add an interaction potential $V_{\rm vv}$ that exists between all pairs of vertices in order to prevent such cases. This potential is defined as
\begin{equation}
    V_{\rm vv} = \sum_{\text{verts }i} \sum_{\text{verts }j} \bar{l}^{\rm vv}_{ij} \varphi(r_{ij}) w_{\rm vv}(t_{ij}, t_{i, j-1}).
\end{equation}
We define this potential as such to enforce continuity with the vertex-edge potential defined above. The potential energy density $\varphi$ is the same as used in the previous section, but now it is a function of the intervertex distance $r_{ij} = \abs{\vb*{r}_j - \vb*{r}_i}$. We define two new functions to help enforce continuity: a particular effective local surface element $\bar{l}^{\rm vv}_{ij}$ and a particular activation function $w_{\rm vv}$. The particular local surface element $\bar{l}_{ij}^{\rm vv}$ for vertex-vertex interactions if defined as
\begin{equation}
    \bar{l}_{ij}^{\rm vv} = \frac{l_i \Xi_{ij-1} + l_{i-1}\Xi_{i-1j}}{2}
\end{equation}
which is similar to the local surface element $\bar{l}_{ij}$ defined above, but now comparison is made between the edge pairs ($i$, $j-1$) and ($i-1$, $j$). The vertex-vertex activation function $w_{\rm vv}$ is designed to only allow interactions when a vertex $i$ is within sufficient distance to another cell surface, but there is no edge upon which the vertex projects (see Fig.~\ref{suppfig:vv_pot_required}). This function is defined as
\begin{equation}
    w_{\rm vv}(t, t') = \begin{cases}
        1 \text{ if } t > 0, \ t' < 1\\
        0 \text{ otherwise }.
    \end{cases}
\end{equation}
This then enforces continuity as a vertex $i$ moves from a ``vertex-edge" interaction zone to a ``vertex-vertex" interaction zone. 

\subsection{Potential Energy Density}\label{suppsec:adhesion:pot_en_density}

The potential energy density considered in this work was introduced Ref.~\cite{Giammona.Campàs.2021}, and in shown in Fig.~\ref{suppfig:pot_en_dens}. It has the form
\begin{equation}\label{suppeq:giammona_campas_pot}
    \varphi(h) = \left( \frac{\epsilon}{\alpha - \beta} \right) S(h) B(h),
\end{equation}
where $\alpha$, $\beta$ and $\epsilon$ are parameters that determine the potential's strength, $B$ is a Lennard-Jones-like interaction function with a hard core and short-range attraction, and $S$ is a support function that smoothly limits the functions interaction range. These functions are defined as 
\begin{subequations}
    \begin{align*}
        S(h) &= \frac{1}{1 + f(h)} \quad ; \quad f(h) = \exp(\frac{h-\sigma}{a})\\
        B(h) &= \frac{\beta + f(\sigma_0)[\beta + (\sigma_0/a)]}{(h / \sigma_0)^\alpha} - \frac{\alpha + f(\sigma_0)[\alpha + (\sigma_0/a)]}{(h / \sigma_0)^\beta}
    \end{align*}
\end{subequations}
Plots of $S$ and $\varphi$ are shown in Fig.~\ref{suppfig:pot_en_dens} for $\alpha = 4$ and $\beta = 3$, which are the values used in this work. The potential has a minimum at $h = \sigma_0$, effective range $\sigma$, cutoff steepness $a$, and curvature determined by $\alpha$ and $\beta$. In this work, we chose $a = 10^{-1}\sigma$, $\sigma_0 = \sigma/2$, and $\sigma$ as $\bar{l}/2$, where $\bar{l}$ is the average polygonal edge length at the start of a simulation. 

\end{document}